\documentclass[12pt,epsf,amssymb]{article} \textheight =23 truecm
\def\lsim{\raise0.3ex\hbox{$<$\kern-0.75em\raise-1.1ex\hbox{$\sim$}}}
\def\gsim{\raise0.3ex\hbox{$>$\kern-0.75em\raise-1.1ex\hbox{$\sim$}}}
\def\noi{\noindent}  \def\bea{\begin{eqnarray}}
\def\eea{\end{eqnarray}} \def\beq{\begin{equation}}
\def\eeq{\end{equation}} 
\def\beeq{\begin{eqnarray}} \def\eeeq{\end{eqnarray}} \def\R{ {\rm R
\kern -.31cm I \kern .15cm}} \def\C{ {\rm C \kern -.15cm \vrule
width.5pt \kern .12cm}} \def\Z{ {\rm Z \kern -.27cm \angle \kern
.02cm}} \def\N{ {\rm N \kern -.26cm \vrule width.4pt \kern .10cm}}
\def\1{{\rm 1\mskip-4.5mu l} }

\usepackage{graphics} \usepackage{graphicx} \usepackage{epsfig}

\begin{document} \begin{center} 

{\large \bf Baryogenesis via leptogenesis from quark-lepton symmetry\par and a compact heavy $N_R$ spectrum}

\par \vskip 1 truecm

{\bf F. Buccella}
\par \vskip 0.5 truemm

{\it Dipartimento di Scienze Fisiche,
Universit\`a di Napoli, Via Cintia, Napoli, Italy}\\    {\it INFN, Sezione di Napoli, Italy}
\par \vskip 2 truemm

{\bf D. Falcone}
\par \vskip 0.5 truemm

{\it Dipartimento di Scienze Fisiche,
Universit\`a di Napoli, Via Cintia, Napoli, Italy}
\par \vskip 2 truemm

{\bf L. Oliver}
\par \vskip 0.5 truemm

{\it Laboratoire de Physique Th\'eorique}\footnote{Unit\'e Mixte de
Recherche UMR 8627 - CNRS }\\    {\it Universit\'e de Paris XI,
B\^atiment 210, 91405 Orsay Cedex, France} 

\end{center}

\begin{abstract} 

By demanding a compact spectrum for the right-handed neutrinos and an approximate quark-lepton symmetry inspired from $SO(10)$ gauge unification (assuming a Dirac neutrino mass matrix close to the up quark mass matrix), we construct a {\it fine tuning} scenario for baryogenesis via leptogenesis. We find two solutions with a normal hierarchy, with the lightest neutrino mass $m_1$ different from zero, providing an absolute scale for the spectrum. In the approximations of the model, there are three independent CP phases : $\delta_L$ (that we take of the order of the quark Kobayashi-Maskawa phase) and the two light neutrino Majorana phases $\alpha$ and $\beta$.
A main conclusion is that, although this general scheme is rather flexible,
in some regions of parameter space we find that the necessary baryogenesis with its sign is given in terms of the $\delta_L$ phase alone. The light Majorana phases can also be computed and turn out to be close of $\pi/2$ or small.
Moreover, $SO(10)$ breaks down to the Pati-Salam group $SU(4) \times SU(2) \times SU(2)$ at the expected natural intermediate scale of about $10^{10}-10^{11}\ GeV$. A prediction is done for the effective mass in $(\beta \beta)_{0\nu}$ decay, the $\nu_e$ mass and the sum of all light neutrino masses.

\end{abstract}

\vskip 0.3 truemm

\noi LPT Orsay 08-65, DSFNA-07-10\par
\noi June 2010

\newpage \pagestyle{plain}

\section{Introduction and qualitative remarks} \hspace*{\parindent} 

The discovery of oscillations, advocated so many years ago by Pontecorvo \cite{P}, in solar and atmospheric neutrinos is one of the most important experimental discoveries of the last century, the most relevant after the proposal of the Standard Model and its precision tests. The discovery of neutrino oscillations is also a milestone in the search of New Physics (NP).

Up to now four quantities related to the Pontecorvo, Maki, Nagakawa and Sakata (PMNS) matrix \cite{MNS}\cite{BP} have been experimentally measured :

\beq
\label{1.1e}
\Delta m_s^2 \simeq 8 \times 10^{-5}\ eV^2
\eeq
\beq
\label{1.2e}
\tan^2 \theta_s \simeq 0.4
\eeq
\beq
\label{1.3e}
\Delta m_a^2 \simeq 2.5 \times 10^{-3}\ eV^2
\eeq
\beq
\label{1.4e}
\tan^2 \theta_a \simeq 1
\eeq
where the subindices $s$ and $a$ mean respectively solar and atmospheric neutrinos.

An upper bound has been been found for the component of $\nu_{eL}$ along the heaviest $\nu_{L}$ mass eigenstate
\beq
\label{1.5e}
\sin^2 \theta_{13} < 0.05
\eeq
and the limits
\beq
\label{1.6e}
m_{\nu_e} < 2.2\ eV
\eeq
\beq
\label{1.7e}
|< m_{ee} >|\ < 0.4\ eV
\eeq
\beq
\label{1.8e}
\sum_i m_{\nu_i} < 1\ eV
\eeq
from the high energy spectrum of the electrons in nuclear beta decay, from the upper limit on the rate in neutrinoless double beta decay (for Majorana neutrinos) and from astrophysics.\par
Interestingly, a more restrictive bound combining all cosmological data has been obtained recently by G. Fogli et al. \cite{FLMMPSSS} :
\beq
\label{1.8bise}
\sum_i m_{\nu_i} < 0.2\ eV
\eeq
to which we will refer in Section 8, comparing it to our results.\par 
But for the moment, in this qualitative introduction, we will rely on the generally accepted loser bound (\ref{1.8e}).\par

The most natural framework to account for the order of magnitude of neutrino masses is the seesaw model \cite{SS}, where the $6 \times 6$ neutrino mass matrix has the form

\beq
\label{1.9e}
\left( \begin{array}{ccc}
0 & m_D^t \\
m_D & M_R
\end{array} \right)
\eeq
where the $3 \times 3$ Dirac neutrino mass matrix $m_D$ has elements of the order of the masses of charged fermions and $M_R$ is the Majorana mass matrix of the right-handed neutrinos, which are singlets of the Standard Model gauge group, with elements of the order of the scale of breaking of the lepton quantum number.\par

The information on oscillations gives us only four of the nine parameters of the light neutrino mass matrix. Within the simplifying assumption of neglecting $\theta_{13}$ and consequently the neutrino Dirac CP violating phase, we will be able to strongly constrain the value of its smallest eigenvalue, and fix the values of the two higher ones, as well as the two Majorana phases, simply by demanding that these parameters have a soft dependence on the values of the matrix elements of $M_R$. We will obtain these results despite the fact that we expect a rather hierarchical spectrum for the eigenvalues of $m_D$, as it happens for the other fermions and as is natural in a $SO(10)$ framework. The mathematical principle is quite simple : it is that the inverse of a function with a critical dependence on a variable is a very slowly varying function : the product of the derivatives is of $O(1)$. The demand of having matrix elements and eigenvalues of $M_R$ of the same order, given a mixing matrix of leptons similar to the one for quarks, will fix $m_1$ and the Majorana phases of light neutrinos. As a result of this requirement, we shall get a compact spectrum for the $N_R$ masses, which will make the leptogenesis scenario for baryogenesis natural, as well as predictions for the electron neutrino mass bounded from tritium $\beta$ decay and the matrix element $|< m_{ee} >|$ appearing in neutrinoless double beta decay. By compact spectrum for the heavy right-handed neutrinos we simply mean to have eigenvalues of the same order of magnitude.\par

From the seesaw formula

\beq
\label{1.10e}
m_L = -m_D\ M_R^{-1}\ m_D^t
\eeq
one gets

\beq
\label{1.11e}
\det M_R = - {(\det m_D)^2 \over \det m_L}
\eeq

From eqn. (\ref{1.8e}) one obtains the upper limit

\beq
\label{1.12e}
|\det m_L| < {1 \over 27}\ eV^3
\eeq
while in principle there is no lower limit for the l.h.s. of the inequality (\ref{1.12e}). Notice that we write the absolute value in the l.h.s. of (\ref{1.12e}) because neutrino masses, being Majorana masses, can differ in sign for neutrinos with opposite CP. \par

Moreover, from eqns. (\ref{1.1e}) and (\ref{1.3e}) we get :
\beq
\label{1.13bise}
\Delta m_s^2 = |m_2|^2 - |m_1|^2  \simeq 8 \times 10^{-5}\ eV^2
\eeq
\beq
\label{1.13e}
\Delta m_a^2 = |m_3|^2 - \textrm{cos}^2\theta_s\ |m_2|^2 - \textrm{sin}^2 \theta_s\ |m_1|^2  \simeq 2.5 \times 10^{-3}\ eV^2
\eeq
where the unfamiliar formula (\ref{1.13e}) for $\Delta m_a^2$, proposed in \cite{BF}, is demonstrated in the Appendix. This formula is an improvement over the usual ones found in the literature, $\Delta m_a^2 = |m_3|^2 - |m_2|^2$ or $\Delta m_a^2 = |m_3|^2 - |m_1|^2$. Of course, in the limit $|m_2| \simeq |m_1|$ all these formulas coincide. However, we must underline that the results of this paper are not really sensitive to adopting formula (\ref{1.13e}) or the usual ones.\par
From the preceding formula one gets a lower limit for the ratio

\beq
\label{1.14e}
{|{m_2 \over m_3}|} > 0.18
\eeq

\par \vskip 0.5 truecm

A temptative lower bound for $|\det m_L|$ may be found in the $SO(10)$ framework by taking, as in \cite{AFS},
\beq
\label{1.14bise}
|\det m_D| = 4 \times 10^{-2}\ GeV^3
\eeq
and for $|\det M_R|$ the upper limit 
\beq
\label{1.14tere}
|\det M_R| \leq 2.7 \times 10^{34}\ GeV^3
\eeq
which comes by assuming that the three right-handed neutrinos take a mass at the scale of $B-L$ spontaneous symmetry breaking in the $SO(10)$ model, with breaking to the $SU(4) \times SU(2) \times SU(2)$ Pati-Salam group \cite{PS} at the intermediate scale $3 \times 10^{11}$ \cite{MOHAPATRA 94}\cite{AABPRT}.\par

We then get, from the seesaw formula (\ref{1.11e}) :
\beq
\label{1.15e}
|\det m_L| \geq 6 \times 10^{-11}\ eV^3
\eeq
Assuming a normal hierarchy for the light neutrinos :
\beq
\label{1.20e}
|m_2| \sim \sqrt{\Delta m_s^2} \simeq 8.9 \times 10^{-3}\ eV 
\eeq
\beq
\label{1.21e}
|m_3| \sim \sqrt{\Delta m_a^2} \simeq 5.0 \times 10^{-2}\ eV
\eeq
eqn. (\ref{1.15e}) will then imply the following lower bound for $|m_1|$ :
\beq
\label{1.21bise}
|m_1| \geq 1.3 \times 10^{-7}\ eV 
\eeq
i.e., {\it a non-vanishing value for the lightest neutrino mass $m_1$}, an absolute scale for the light neutrino spectrum.\par 

As we will see below, a rather sharp prediction for $m_1$ and relevant predictions for the l.h.s. of eqns. (\ref{1.6e})-(\ref{1.8e}) will be achieved by our demand of a compact $M_R$ spectrum and successful leptogenesis.

\par \vskip 0.5 truecm

The measurements of the cosmic microwave background anisotropies \cite{WMAP} and the abundance of light nuclei produced in primordial nucleosynthesis \cite{BBN} give a consistent value for the baryon asymmetry :

\beq
\label{1.22e}
Y_B = {n_B-n_{\bar{B}} \over s} \simeq {1 \over 7.04} {n_B-n_{\bar{B}} \over n_\gamma} \simeq 9 \times 10^{-11} 
\eeq

This baryonic asymmetry may arise from the leptogenesis scenario \cite{LEPTOGENESIS}, with a leptonic asymmetry produced at a high scale, which gives rise by the $B-L$ conserving sphaleron processes \cite{KRS} at the electroweak scale to a baryon asymmetry below that scale. 

Within the leptogenesis scenario, the baryon asymmetry, baryon to entropy fraction, is given by

\beq
\label{1.23e}
Y_B \simeq - \frac{1}{2} Y_L
\eeq
that should be compared with the experimental value given by (\ref{1.22e}). 

Concerning Grand Unification, the $SU(5)$ minimal model is disfavored, since it generates a small baryon asymmetry at the high scale, washed out at the electroweak scale, since in that model $B-L$ is conserved. Thus, $SO(10)$ with its $B-L$ generator spontaneously broken, that we will adopt in its non-Supersymmetric version, should be preferred to $SU(5)$ to realize the leptogenesis scenario.\par

The paper is organized as follows. In Section 2 we give the relevant formulas for the inverse seesaw, mass matrices and mixings. In Section 3 we formulate our $SO(10)$ Ansatz. In Section 4 we give the formulas needed for $CP$ violation and the baryon asymmetry. Section 5 is devoted to a simple mathematical procedure to obtain a quasi-degenerate right-handed neutrino spectrum (that presents a level crossing) and a realistic light neutrino spectrum. We underline an illuminating limit of considering, for the matrix diagonalizing $m_D$, a pure Cabibbo matrix that then we extend to a general matrix of the CKM form, introducing therefore CP violation. We find two possible solutions. In Section 6 we expose a simple procedure to slowly lift the degeneracy of the heavy right handed neutrinos, and give the corresponding evolution of $\Delta m_s^2$ and $\Delta m_a^2$. In Section 7 we exhibit the results for CP violation and baryon asymmetry, in the one-flavor approximation, and in Section 8 we give the predictions for $m_{\nu_e}$ and the effective neutrino mass in $(\beta \beta)_{0\nu}$. In Section 9 we relax a reality assumption used in Sections 6 and 7. In Section 10 we comment on the compact heavy neutrino spectrum and on the level crossing region. Finally in Sections 11 and 12 we underline open problems within the present approach and we conclude. 

\section{Inverse seesaw, mass matrices and mixings} \hspace*{\parindent} 

From (\ref{1.10e}), we can deduce the inverse seesaw formula, 

\beq
\label{2.1e}
M_R = -m_D^t\ m_L^{-1}\ m_D
\eeq
and diagonalizing the neutrino Dirac mass matrix $m_D$ by

\beq
\label{2.2e}
m_D = V^{L+}\ m_D^{diag}\ V^R
\eeq
one gets the formula \cite{BF}

$$M_R = -m_D^t\ m_L^{-1}\ m_D = -\ V^{Rt}\ m_D^{diag}\ V^{L*}\ m_L^{-1}\ V^{L+}\ m_D^{diag}\ V^R$$ 
\beq
\label{2.3e}
= -\ V^{Rt}\ m_D^{diag}\ A^L\ m_D^{diag}\ V^R \qquad \qquad \qquad \qquad \qquad \ \ \ 
\eeq
where the last equality follows from the definition \cite{BF} of the matrix $A^L$

\beq
\label{2.4e}
A^L = V^{L*}\ m_L^{-1}\ V^{L+}
\eeq

\noindent The neutrino mass matrix $m_L$ is diagonalized by the PMNS matrix $U$ :

\beq
\label{2.5e}
m_L = U^*\ m_L^{diag}\ U^+
\eeq
where 

\beq
\label{2.5bise}
m_L^{diag} = \textrm{diag}(m_1,m_2,m_3)
\eeq
and $U$ writes :

$$ U = \left(
        \begin{array}{ccc}
        c_{12}c_{13} & s_{12}c_{13} & s_{13} e^{-i\delta} \\
        -s_{12}c_{23}-c_{12}s_{23}s_{13} e^{i\delta} & c_{12}c_{23}-s_{12}s_{23}s_{13} e^{i\delta} & s_{23}c_{13} \\
        s_{12}s_{23}-c_{12}c_{23}s_{13} e^{i\delta} & -c_{12}s_{23}-s_{12}c_{23}s_{13} e^{i\delta} & c_{23}c_{13} \\
        \end{array} \right) $$
\beq
\label{2.6e} .\ \textrm{diag}(1,e^{i\alpha},e^{i\beta})
\eeq
where $\delta$ is the Dirac phase and $\alpha$ and $\beta$ are the Majorana phases. For the latter we adopt the convention of Davidson et al. \cite{LEPTOGENESIS}.\par

Let us make a remark on the counting of  phases. One has, in all generality, 6 independent phases in the Type I seesaw scheme, as established in \cite{SANTAMARIA}\cite{EMOP} and as it is exposed in the review \cite{LEPTOGENESIS} (last reference, Section 2.1). In the model that we develop below, the number of independent phases will be reduced according to the hypotheses adopted.

Taking into account the data on solar and atmosperic neutrinos and the fact that $s_{13}$ is bounded to be small, we will take

\beq
\label{2.5tere}
s_{13} \simeq 0
\eeq 
and approximate, from now on, the matrix $U$ as follows :

\beq
\label{2.6-1e}
U \simeq \left(
        \begin{array}{ccc}
       c_s & s_s & 0 \\
        -\ {s_s \over \sqrt{2}} & {c_s \over \sqrt{2}} & {1 \over \sqrt{2}} \\
        {s_s \over \sqrt{2}} & -\ {c_s \over \sqrt{2}} & {1 \over \sqrt{2}} \\
        \end{array}
        \right).\ \textrm{diag}(1,e^{i\alpha},e^{i\beta})
\eeq
and CP violation originating in the Dirac phase $\delta$ drops out.
 
To simplify the expressions in what follows, we change the notation for the diagonal matrix in (\ref{2.5e})-(\ref{2.6e})
 
\beq
\label{2.6-2e}
\textrm{diag}(m_1,e^{-2i\alpha}m_2,e^{-2i\beta}m_3) \qquad \to \qquad \textrm{diag}(m_1,m_2,m_3)
\eeq
where, from now on, $m_2$ and $m_3$ are assumed to be {\it complex parameters}.\par

Eqn. (\ref{2.5e}) now writes, in the approximation (\ref{2.6-1e}), and with the notation convention of the r.h.s. of (\ref{2.6-2e}) :

\beq
\label{2.6-3e}
m_L \simeq \left(
        \begin{array}{ccc}
       c_s & s_s & 0 \\
        -\ {s_s \over \sqrt{2}} & {c_s \over \sqrt{2}} & {1 \over \sqrt{2}} \\
        {s_s \over \sqrt{2}} & -\ {c_s \over \sqrt{2}} & {1 \over \sqrt{2}} \\
        \end{array}\right)\ \textrm{diag}(m_1,m_2,m_3)\ \left(
        \begin{array}{ccc}
       c_s & -\ {s_s \over \sqrt{2}} & {s_s \over \sqrt{2}} \\
         s_s & {c_s \over \sqrt{2}} & -\ {c_s \over \sqrt{2}} \\
        0 & {1 \over \sqrt{2}} & {1 \over \sqrt{2}} \\
        \end{array}\right)
\eeq

Let us strongly underline again that in (\ref{2.6-3e}), and in what follows, the parameters $m_2, m_3$ are assumed to be complex, containing, according to (\ref{2.6-2e}), the Majorana phases defined by (\ref{2.6e}).\par 
These phases will be computed at different stages. Their calculation will depend on some hypotheses to be made explicit below, and on the successive parametrizations assumed for the matrix $m_D$ (\ref{2.2e}).\par 

From the previous hypotheses we obtain the following complex symmetric matrix
 
\beq
\label{2.7e}
m_L^{-1} =  \left(
        \begin{array}{ccc}
        {c_s^2 \over m_1} + {s_s^2 \over m_2} & - {c_s s_s \over \sqrt{2}} \left ({1 \over m_1} - {1 \over m_2} \right ) & {c_s s_s \over \sqrt{2}} \left ({1 \over m_1} - {1 \over m_2} \right ) \\  - {c_s s_s \over \sqrt{2}} \left ({1 \over m_1} - {1 \over m_2} \right )  & {1 \over 2}  \left ({s_s^2 \over m_1} + {c_s^2 \over m_2} + {1 \over m_3} \right ) &  - {1 \over 2}  \left ({s_s^2 \over m_1} + {c_s^2 \over m_2} - {1 \over m_3} \right ) \\   {c_s s_s \over \sqrt{2}} \left ({1 \over m_1} - {1 \over m_2} \right ) &  - {1 \over 2}  \left ({s_s^2 \over m_1} + {c_s^2 \over m_2} - {1 \over m_3} \right ) &  {1 \over 2}  \left ({s_s^2 \over m_1} + {c_s^2 \over m_2} + {1 \over m_3} \right ) \\
        \end{array}
        \right)
 \eeq
 and the matrix $A_L$ (\ref{2.4e}) is also complex symmetric : 
 
\beq
\label{2.7bise}
A^{Lt} = A^{L} 
\eeq

In a previous work \cite{BF}, to comply with the lower bound for the mass of the lightest right-handed neutrino claimed by \cite{DI}, we did set upper limits on the coefficients of contributions proportional to the products of the Dirac matrix eigenvalues $(m_{D_3})^2$ and
$m_{D_2} m_{D_3}$ in the $M_R$ matrix, related to $m_L$ by the inverse seesaw formula.\par

From formula (\ref{2.3e}) we see that to get a quasi-degenerate heavy Majorana neutrino spectrum we need that the terms proportional to $(m_{D_3})^2$ and $m_{D_2} m_{D_3}$ have to be small, that means for a hierarchical Dirac mass spectrum that the matrix elements of (\ref{2.4e}) $A^L_{33}$ and $A^L_{23} = A^L_{32}$ have to be small.
From (\ref{2.4e}), we get the following expression for these matrix elements of interest :

$$A^L_{23} = V^{L*}_{31} \left [ \left ({c_s^2 \over m_1} + {s_s^2 \over m_2} \right )V^{L*}_{21} - {c_s s_s \over \sqrt{2}} \left ({1 \over m_1} - {1 \over m_2} \right )V^{L*}_{22} + {c_s s_s \over \sqrt{2}} \left ({1 \over m_1} - {1 \over m_2} \right )V^{L*}_{23} \right ]$$
$$+ V^{L*}_{32} \left [ - {c_s s_s \over \sqrt{2}} \left ({1 \over m_1} - {1 \over m_2} \right )V^{L*}_{21} +  {1 \over 2}  \left ({s_s^2 \over m_1} + {c_s^2 \over m_2} + {1 \over m_3} \right )V^{L*}_{22} - {1 \over 2}  \left ({s_s^2 \over m_1} + {c_s^2 \over m_2} - {1 \over m_3} \right )V^{L*}_{23} \right ]$$ 
 \beq
\label{2.8e}
+ V^{L*}_{33} \left [ {c_s s_s \over \sqrt{2}} \left ({1 \over m_1} - {1 \over m_2} \right )V^{L*}_{21} -  {1 \over 2}  \left ({s_s^2 \over m_1} + {c_s^2 \over m_2} - {1 \over m_3} \right )V^{L*}_{22} + {1 \over 2}  \left ({s_s^2 \over m_1} + {c_s^2 \over m_2} + {1 \over m_3} \right )V^{L*}_{23} \right ]
\eeq

$$A^L_{33} = V^{L*}_{31} \left [ \left ({c_s^2 \over m_1} + {s_s^2 \over m_2} \right )V^{L*}_{31} - {c_s s_s \over \sqrt{2}} \left ({1 \over m_1} - {1 \over m_2} \right )V^{L*}_{32} + {c_s s_s \over \sqrt{2}} \left ({1 \over m_1} - {1 \over m_2} \right )V^{L*}_{33} \right ]$$
$$+ V^{L*}_{32} \left [ - {c_s s_s \over \sqrt{2}} \left ({1 \over m_1} - {1 \over m_2} \right )V^{L*}_{31} +  {1 \over 2}  \left ({s_s^2 \over m_1} + {c_s^2 \over m_2} + {1 \over m_3} \right )V^{L*}_{32} - {1 \over 2}  \left ({s_s^2 \over m_1} + {c_s^2 \over m_2} - {1 \over m_3} \right )V^{L*}_{33} \right ]$$ 
\beq
\label{2.9e}
+ V^{L*}_{33} \left [ {c_s s_s \over \sqrt{2}} \left ({1 \over m_1} - {1 \over m_2} \right )V^{L*}_{31} -  {1 \over 2}  \left ({s_s^2 \over m_1} + {c_s^2 \over m_2} - {1 \over m_3} \right )V^{L*}_{32} + {1 \over 2}  \left ({s_s^2 \over m_1} + {c_s^2 \over m_2} + {1 \over m_3} \right )V^{L*}_{33} \right ]
\eeq

\section{Our $SO(10)$ Ansatz} \hspace*{\parindent} 

The seesaw model is realized in the framework of $SO(10)$ unified gauge theories \cite{SO10}, where $B-L$ is a generator, which has to be spontaneously broken. Long time before the firm evidence for neutrino oscillations this phenomenon had been claimed \cite{BCST} as the most promising experimental signal for $SO(10)$ unification.\par

A systematic study of the spontaneous symmetry breaking in $SO(10)$ unified theories has lead to propose \cite{MOHAPATRA 94} the model with $SU(4) \times SU(2) \times SU(2)$ \cite{PS} intermediate gauge group, broken at the scale of order $3 \times 10^{11}\ GeV$ \cite{AABPRT}.\par

A general analysis has been done in \cite{AFS} on the possibility to construct a realistic leptogenesis scenario within the seesaw model with neutrino Dirac masses in a hierarchical ratio, as it is the case for u-type quarks. The most promising case has been found with $M_3 \sim 10^{14}\ GeV$ and nearby values for the masses of the two lightest right-handed neutrinos.\par 

Although in the present paper we follow the general idea \cite{AFS} of leptogenesis generated by quasi-degenerate right-handed neutrinos, we look for a more compact spectrum for $N_R$, with the heaviest right-handed neutrino at the intermediate scale, of the order $10^{11}\ GeV$.\par

In $SO(10)$ the hypothesis that the electroweak Higgs transforms as a combination of {\bf 10} representations implies at the unification scale the equalities among mass matrices

\beq
\label{3.1e}
m_e = m_d
\eeq
\beq
\label{3.2e}
m_D = m_u
\eeq

For $b$ and $\tau$ masses relation (\ref{3.1e}) at the intermediate scale is in reasonable agreement with experiment but, as Georgi and Jarlskog \cite{GJ} have shown in the $SU(5)$ case, one needs also higher dimensional representations. The generalization of this argument to $SO(10)$ was given by Harvey et al. \cite{HRR}. For an overview on fermion masses and mixings in gauge theories, see the review article \cite{FALCONE}.

Within $SO(10)$, with the electroweak Higgs boson belonging to the {\bf 10} and/or {\bf 126} representations, and {\it no component along the {\bf 120} representation}, the mass matrices are symmetric. As a consequence, the unitary matrices $V^R$ and $V^L$ that diagonalize Dirac neutrino matrix (\ref{2.2e}) are related :

\beq
\label{3.3e}
V^R = V^{L*}
\eeq
and the matrix $M_R$ (\ref{2.3e}) becomes 

$$M_R = -m_D^t\ m_L^{-1}\ m_D = -\ V^{L+}\ m_D^{diag}\ V^{L*}\ m_L^{-1}\ V^{L+}\ m_D^{diag}\ V^{L*}$$
\beq
\label{3.4e}
= -\ V^{L+}\ m_D^{diag}\ A^L\ m_D^{diag}\ V^{L*} \qquad \qquad \qquad \qquad \qquad \ \ \
\eeq

Let us now go back to the question of the phase counting, quoted for the seesaw scheme in Section 3, in the particular case of $SO(10)$ with symmetric Dirac neutrino matrix. Since $V^L$ has only one phase, and $m_L$, through the mixing matrix $U$ (\ref{2.6e}) has three phases, we have reduced the number of independent phases, from 6 in the general case to 4 independent phases. In the approximation (\ref{2.5tere}) $s_{13} \simeq 0$ that we have adopted, this means that we have 3 independent phases, namely a phase from $V^L$, that we will call $\delta_L$, and the two Majorana phases $\alpha$ and $\beta$ from (\ref{2.6-1e}). \par
Below, in Sections 5 and 6, we will impose two other conditions that reduce further the number of independent phases, from 3 to a single one.\par

For the diagonalized Dirac neutrino matrix

\beq
\label{3.6e}
m_D^{diag} = \left(
        \begin{array}{ccc}
       m_{D_1} & 0 & 0 \\
        0 & m_{D_2} & 0 \\
        0 & 0 & m_{D_3} \\
        \end{array}
        \right)
\eeq
we will adopt the numerical values proposed in \cite{AFS}, inspired from the up-quark mass matrix :
\beq
\label{3.7e}
m_{D_1} = 10^{-3}\ GeV \qquad \qquad m_{D_2} = 0.4\ GeV \qquad \qquad m_{D_3} = 100\ GeV
\eeq

The matrix $m_D^{diag}\ A^L\ m_D^{diag}$ appearing in (\ref{2.3e}) has the form
\beq
\label{3.8e}
m_D^{diag}\ A^L\ m_D^{diag} = \left(
        \begin{array}{ccc}
       m_{D_1}^2\ A^L_{11} & m_{D_1} m_{D_2}\ A^L_{12} & m_{D_1} m_{D_3}\ A^L_{13} \\
        m_{D_1} m_{D_2}\ A^L_{12} &  m_{D_2}^2\ A^L_{22} & m_{D_2} m_{D_3}\ A^L_{23} \\
        m_{D_1} m_{D_3}\ A^L_{13} & m_{D_2} m_{D_3}\ A^L_{23} & m_{D_3}^2\ A^L_{33} \\
        \end{array}
        \right)
\eeq
that clearly shows that in order to have a compact $N_R$ spectrum from (\ref{2.3e}) one needs small values for the matrix elements $A^L_{23}$ and $A^L_{33}$.
 
For $V^L$ we will assume a form qualitatively similar to the Cabibbo-Kobayashi-Maskawa (CKM) quark matrix, that reads, in the standard convention (except for the phase $\delta_L$, we take the same notation as for the light neutrino mixing matrix (\ref{2.6e}), but in what follows there is no ambiguity) :

\beq
\label{3.9e}
V^L = \left(
        \begin{array}{ccc}
        c_{12}c_{13} & s_{12}c_{13} & s_{13} e^{-i\delta_L} \\
        -s_{12}c_{23}-c_{12}s_{23}s_{13} e^{i\delta_L} & c_{12}c_{23}-s_{12}s_{23}s_{13} e^{i\delta_L} & s_{23}c_{13} \\
        s_{12}s_{23}-c_{12}c_{23}s_{13} e^{i\delta_L} & -c_{12}s_{23}-s_{12}c_{23}s_{13} e^{i\delta_L} & c_{23}c_{13} \\
        \end{array}
        \right)
\eeq
where $\delta_L$ is the CP-violating phase.\par

Formula (\ref{3.9e}) is correct at least for exact quark-lepton symmetry, with Higgs in the
{\bf 10} representation : if the mass matrices (\ref{3.1e}) are diagonal and real one has $V^L = V_{CKM}$. For phenomenological purposes we assume this form in what follows.\par

We define, as usual, in terms of Wolfenstein parameters :
\beq
\label{3.9bise}
s_{12} = \lambda \qquad \qquad s_{23} = A\lambda^2 \qquad \qquad s_{13}e^{i\delta_L} = A\lambda^3(\rho+i\eta)
\eeq
Of course, in our problem the parameters $\lambda, A, \rho, \eta$ do not necessarily have the same precise values as in the quark sector : we are interested only in an order of magnitude estimate.  

Let us say some words concerning the diagonalization of the right-handed neutrino matrix. Since in our $SO(10)$ Ansatz $M_R$ (\ref{3.4e}) is complex and symmetric, we can diagonalize it by using a single unitary matrix :

\beq
\label{3.10e}
M_R = W_R\ M_R^{diag}\ W_R^t
\eeq

The matrix $W_R$ is such that all eigenvalues are real and positive. The effect of phases will appear in the matrix $W_R$. These phases will of course have consequences for baryogenesis and for neutrinoless double beta decay.\par

As we  will see below, our demand of suppressed values for $A^L_{33}$ and $A^L_{23}$ generates a compact form for the $M_R$ spectrum, which helps in getting in a natural way the desired lepton asymmetry. 

\par \vskip 0.8 truecm

\section{Leptogenesis and baryon asymmetry} \hspace*{\parindent} 

In this Section we recall the basic formulas concerning the CP violating asymmetry $\epsilon_1$ and the corresponding baryogenesis asymmetry $Y_{B_1}$. We work in the basis in which the mass matrices of charged leptons and of right-handed neutrinos are diagonal, i.e. from (\ref{2.1e}) and (\ref{3.10e}) :

\beq
\label{4.1e}
M_R^{diag} = - W_R^+\ m_D^t\ m_L^{-1}\ m_D\ W_R^* 
\eeq

\noindent Therefore, in the computation of the CP-violating asymmetry $\epsilon_1$ we define

\beq
\label{4.2e}
\hat{m}_D = m_D\ W_R^*
\eeq
such that 

\beq
\label{4.2ae}
M_R^{diag} = - \hat{m}_D^t\ m_L^{-1}\ \hat{m}_D
\eeq
By convention we label the masses of the heavy neutrinos $N_{R_i}$ (i = 1, 2, 3) :
\beq
\label{4.2bise}
0 \leq M_1 \leq M_2 \leq M_3
\eeq

\noindent In terms of $\hat{m}_D$, the CP asymmetry writes, for the lightest heavy neutrino $N_{R_1}$ :

\beq
\label{4.3e}
\epsilon_1 = {1 \over 8\pi v^2} \sum_{k\neq 1} f\left({M_k^2 \over M_1^2}\right) {\textrm{Im}\left[(\hat{m}_D^+\hat{m}_D)_{1k}^2\right] \over (\hat{m}_D^+\hat{m}_D)_{11}}
\eeq
where $v = 174\ GeV$ is the scale of electroweak symmetry breaking, and the function $f(x)$ is given by \cite{FUNCTION} :

\beq
\label{4.4e}
f(x) = \sqrt{x} \left[{1 \over 1-x} +1 - (1+x)\ \log \left({1+x \over x}\right)\right]
\eeq
that in the limit $x >> 1$ becomes :
\beq
\label{4.5e}
f(x) \simeq -{3 \over 2\sqrt{x}}
\eeq
and the effective neutrino mass, that controls the amount of washout, writes :

\beq
\label{4.6e}
{\tilde m}_1 = {(\hat{m}_D^+\hat{m}_D)_{11} \over M_1}
\eeq

\noindent The cases that we encounter in our calculations below satisfy the strong washout condition

\beq
\label{4.7e}
{\tilde m}_1 >> 3 \times 10^{-3}\ eV
\eeq
and the corresponding baryon asymmetry writes, {\it in the one-flavor approximation}, that we will adopt in the following \cite{WASHOUT} :

\beq
\label{4.8e}
Y_{B_1} = - {1 \over 2}\ 0.3\ {\epsilon_1 \over g_*}\ \left({0.55 \times 10^{-3} eV \over {\tilde m}_1}\right)^{1.16}
\eeq
where $g_* \simeq 107$ in the Standard Model, in the non-Supersymmetric case.

\par \vskip 1.0 truecm

\section{Quasi-degenerate heavy right-handed neutrinos and realistic light neutrino spectrum} \hspace*{\parindent} 

In order to get a compact $N_R$ spectrum, a sufficient condition is to impose that the matrix elements $A^L_{33}$ and $A^L_{23}$ are suppressed, because we are dealing with the matrix (\ref{3.8e}) and the $m_D$ eigenvalues (\ref{3.7e}). As a first exercise, we thus consider the solutions of the equations, linear and homogeneous in the inverse of the neutrino masses ${1 \over m_i}\ (i = 1, 2, 3)$,

\beq
\label{5.1e}
A^L_{23}(m_1,m_2,m_3) = A^L_{33}(m_1,m_2,m_3) = 0 
\eeq
We are aware that this is a very drastic assumption, but will help to guide our research of a compact right-handed neutrino spectrum, and also to look for its consequences on the light neutrino masses and the amount of baryogenesis that one can get. We must emphasize that in this Section, and in the following ones, we are dealing with a {\it fine tuning} scheme. We cannot content ourselves with just order-of-magnitude estimates, but we need precise numerical calculations.\par
Notice a new important point in the phase counting of eqn. (\ref{3.4e}) with the hypothesis (\ref{2.5tere}). Under the two reality conditions (\ref{5.1e}) that we now impose, the 3 phases (see Section 3) are now reduced to a single phase, either $\delta_L$ or one of the two Majorana phases $\alpha$ or $\beta$.\par 
Since we do not have experimental information on the Majorana phases, we will, from now on, compute $\alpha$ and $\beta$, and later the CP asymmetry $\epsilon_1$ and baryon asymmetry $Y_{B_1}$ in terms of $\delta_L$. Of course, in principle one could also compute the pair ($\delta_L$, $\alpha$) in terms of $\beta$ or ($\delta_L$, $\beta$) in terms of $\alpha$. But in the present $SO(10)$ approach the natural thing to do is to take as input $\delta_L$, since we can take it to be of the order of Kobayashi-Maskawa (KM) phase $\delta_{KM}$, on which we have information.

\subsection{$V^L$ in the limit of a pure Cabibbo matrix}

\par \vskip 0.5 truecm

For our purpose, it is a good illustration to study the consequences of this hypothesis considering it within the very simplified approximation of a $2 \times 2$ Cabibbo matrix 

\beq
\label{5.1bise}
V^L = \left(
        \begin{array}{ccc}
        c_{12} & s_{12} & 0 \\
        -s_{12} & c_{12} & 0 \\
        0 & 0 & 1 \\
        \end{array}
        \right)
\eeq
\par
Since from (\ref{5.1bise}) $\delta_L$ drops out, we are left with only two phases, namely the Majorana phases $\alpha$ and $\beta$. Imposing the two reality conditions (\ref{5.1e}), these phases will be fixed, as we see below.\par

From (\ref{5.1e}) and (\ref{5.1bise}), we find that the light and heavy neutrino spectra turn out to be reasonable. From (\ref{2.8e})(\ref{2.9e}), the matrix elements of $A^L$ we are interested in are
$$A^L_{23} = A^L_{32} = - {1 \over 2} \left({s_s^2 \over m_1} + {c_s^2 \over m_2} - {1 \over m_3} \right)c_{12}\ -\ {c_s s_s \over \sqrt{2}} \left({1 \over m_1} - {1 \over m_2} \right)s_{12}\qquad \qquad \qquad \qquad \qquad$$
\beq
\label{5.8e}
A^L_{33} = {1 \over 2} \left({s_s^2 \over m_1} + {c_s^2 \over m_2} + {1 \over m_3} \right) \qquad \qquad \qquad \qquad \qquad \qquad \qquad \qquad \qquad \qquad 
\eeq
If we impose the very strong assumption (\ref{5.1e}) we have the two equations
$$\qquad \qquad  \qquad {s_s^2 \over m_1} + {c_s^2 \over m_2} - {1 \over m_3} + \sqrt{2} c_s s_s \left({1 \over m_1} - {1 \over m_2} \right) \tan\theta_{12} = 0 \qquad \qquad \qquad \qquad \qquad \qquad$$
\beq
\label{5.9e}
 \qquad \qquad \qquad {s_s^2 \over m_1} + {c_s^2 \over m_2} + {1 \over m_3} = 0  \qquad \qquad \qquad \qquad \qquad \qquad \qquad \qquad \qquad \qquad \qquad
\eeq
and solving for $m_2$ and $m_3$ in terms of $m_1$, one gets :

$$m_2 = - {\sqrt{2} - \textrm{tan}\theta_s\tan\theta_{12} \over \sqrt{2}\ \textrm{tan}^2\theta_s + \textrm{tan}\theta_s\tan\theta_{12}}\ m_1$$
\beq
\label{5.10e}
m_3 = {\sqrt{2} - \textrm{tan}\theta_s\tan\theta_{12} \over {\textrm{tan}\theta_s\tan\theta_{12}}} \ m_1
\eeq

\noindent From the data (\ref{1.1e})-(\ref{1.4e}) and formulas (\ref{1.13bise})(\ref{1.13e}) one gets, from (\ref{5.10e}), a number of solutions for $m_1$ and $\theta_{12}$.\par 

However, if one looks for solutions with $\theta_{12}$ {\it in the neighborhood of the Cabibbo angle} $\theta_C$, one finds the following solutions, according to the sign of $\tan \theta_s$, since only its square (\ref{1.2e}) is measured :

\par \vskip 0.3 truecm

(1) For $\tan \theta_s \simeq - \sqrt{0.4}$ one gets,

\beq
\label{5.11previouse}
\tan \theta_{12} = 0.140
\eeq
\beq
\label{5.11e}
|m_1| = 0.0030\ eV \qquad \qquad m_2 = -3.1522\ m_1 \qquad \qquad m_3 = -16.9273\ m_1
\eeq

We have taken several digits for the $m_i$ values to get a compact spectrum for the $N_{R_i}$ and, as we will see below, for later to obtain the non-degeneracy of the two higher states. The reason is that we are dealing with a fine-tuning problem. Of course, one could take the first digits and say the degree of approximation at each stage, but we believe that our way of presenting the results, although harder to read, corresponds better to the reality of the calculation.\par

In consistency with (\ref{2.6-2e}) we assume the convention $m_1 > 0$ and one gets the following spectrum
\beq
\label{5.12e}
m_1 = 0.0030\ eV \qquad \qquad m_2 = -0.0094\ eV \qquad \qquad m_3 = -0.0507\ eV 
\eeq
where two heavier neutrinos have opposite CP from the lighter one. This means that the Majorana phases are 

\beq
\label{5.12bise}
\alpha = {\pi \over 2} \qquad \qquad \qquad \beta = {\pi \over 2}
\eeq

\par
\noindent We find, from (\ref{2.2e}), (\ref{3.3e}) and the value $\tan\theta_{12}$ (\ref{5.11previouse}), for the symmetric Dirac mass matrix :

\beq
\label{5.13e}
m_D = \left(
        \begin{array}{ccc}
        0.0087 & -0.0549 & 0 \\
        -0.0549 & 0.3923 & 0 \\
        0 & 0 & 100 \\
        \end{array}
        \right)\ GeV
\eeq
and 
\beq
\label{5.14e}
M_1 = 5.5504 \times 10^9\ GeV \qquad M_2 = 1.42991 \times 10^{10}\ GeV \qquad M_3 = 1.42992 \times 10^{10}\ GeV
\eeq

(2) For $\tan \theta_s \simeq + \sqrt{0.4}$ one gets 

\beq
\label{5.15previouse}
\tan \theta_{12} = 0.243
\eeq
\beq
\label{5.15e}
|m_1| = 0.0062\ eV \qquad \qquad m_2 = -1.752\ m_1 \qquad \qquad m_3 = 8.191\ m_1
\eeq

\noindent Assuming again the convention $m_1 > 0$ one gets the following hierarchical spectrum
\beq
\label{5.16e}
m_1 = 0.0062\ eV \qquad \qquad m_2 = -0.0109\ eV \qquad \qquad m_3 = 0.0509\ eV 
\eeq
where two neutrinos (the lightest and the heaviest) have opposite CP from the third one. This means that the Majorana phases are 

\beq
\label{5.16bise}
\alpha = - {\pi \over 2} \qquad \qquad \qquad \beta = 0
\eeq

\par \vskip 0.3 truecm

\noindent We obtain for this solution, from the value $\tan\theta_{12}$ (\ref{5.15previouse}), the Dirac neutrino matrix 

\beq
\label{5.17e}
m_D = \left(
        \begin{array}{ccc}
        0.0233 & -0.0916 & 0 \\
        -0.0916 & 0.3777 & 0 \\
        0 & 0 & 100 \\
        \end{array}
        \right)\ GeV
\eeq
and the quasi-degenerate right-handed heavy neutrino spectrum :
\beq
\label{5.18e}
M_1 = 6.72168 \times 10^9\ GeV \qquad M_2 = 8.30366 \times 10^9\ GeV \qquad M_3 = 8.30409 \times 10^9\ GeV
\eeq

\noindent The results for these two solutions seem encouraging because we get in both cases a value of the angle $\theta_{12}$ that is rather close to the Cabibbo angle $\theta_C$. It seems highly non-trivial and amazing that such a simplified form of the $V^L$ matrix could give already these results consistent with quark-lepton symmetry.

\par \vskip 0.5 truecm

\subsection{$V^L$ with approximate CKM form}

\par \vskip 0.5 truecm

We now switch on the other $V^L$ parameters and consider the full matrix (\ref{3.9e}).
To perform the calculations we adopt for $m_1$ and $\tan \theta_{12}$ the values obtained in the pure Cabibbo limit, namely $m_1 = 0.0030\ eV$, $\tan \theta_{12} = 0.140$ for solution (1) and $m_1 = 0.0062\ eV$, $\tan \theta_{12} = 0.243$ for solution (2). \par 
For the numerical calculations we thus proceed in the following way :

\par \vskip 0.1 truecm

i) From either solution (1) or solution (2) of the preceding Subsection, we fix the parameters $m_1$ and $\tan\theta_{12}$ :
\beq
\label{5.19e}
(1) \qquad \tan\theta_s = -\sqrt{0.4} \qquad \qquad m_1 = 0.0030\ eV \qquad \qquad \tan \theta_{12} = 0.140
\eeq
\beq
\label{5.20e}
(2) \qquad \tan\theta_s = +\sqrt{0.4} \qquad \qquad m_1 = 0.0062\ eV \qquad \qquad \tan \theta_{12} = 0.243
\eeq 

\par \vskip 0.1 truecm
ii) For the rest of the $V^L$ matrix elements we deduce the Wolfenstein parameter $\lambda$ in (\ref{3.9bise}) from (\ref{5.19e}) and (\ref{5.20e}) and take, just a guess, the parameters $A$, $\rho$ and $\eta$ from the quark sector CKM matrix, i.e. for example 
\beq
\label{5.20bise}
A = 0.8 \qquad \qquad \qquad \rho = 0.13 \qquad \qquad \qquad \eta = 0.35 
\eeq
and, using these parameters, we fix $s_{23}, s_{13}$ following (\ref{3.9bise}) :
\beq
\label{5.21e}
(1) \qquad \tan \theta_{12} = 0.140, \qquad s_{23} = 0.0154, \qquad s_{13} = 0.0008, \qquad \delta_L = 1.2152
\eeq 
gives
\beq
\label{5.21bise} 
V^L \simeq \left(
        \begin{array}{ccc}
       0.990 & 0.139 & (2.8-7.5i) \times 10^{-4} \\
        -0.139 & 0.990 & 0.015 \\
        (18.6-7.4i) \times 10^{-4} & -0.015 & 1. \\
        \end{array}
        \right)
\eeq
while from 
\beq
\label{5.22e}
(2) \qquad \tan \theta_{12} = 0.243, \qquad s_{23} = 0.0446, \qquad s_{13} = 0.0039, \qquad \delta_L = 1.2152
\eeq
one obtains
\beq
\label{5.22bise} 
V^L \simeq \left(
        \begin{array}{ccc}
       0.972 & 0.236 & (1.37-3.68i) \times 10^{-3} \\
        -0.236 & 0.971 & 0.045 \\
        (9.20-3.58i) \times 10^{-3} & -0.044 & 1. \\
        \end{array}
        \right)
\eeq

\par \vskip 0.1 truecm

iii) Then, we solve equations (\ref{5.1e}) for $m_2$ and $m_3$ and compare with experiment for $\Delta m_s^2$ and $\Delta m_a^2$.

\par \vskip 0.5 truecm

Notice that this numerical procedure is less rigid that the one adopted in the simpler case of a pure Cabibbo matrix of the preceding Subsection, where $\Delta m_s^2$ and $\Delta m_a^2$ were fixed to the experimental central values (\ref{1.1e})(\ref{1.3e}) and we did solve for $\tan \theta_{12}$, $m_1$, $m_2$ and $m_3$. We prefer to change here our numerical approach due to the extreme fine tuning of the problem. It seems to us sensible enough if we get results for $\Delta m_s^2$ and $\Delta m_a^2$ that are roughly consistent with experiment.

\par \vskip 0.5 truecm

We find the following results.

\par \vskip 0.5 truecm

(1) For $\tan \theta_s \simeq - \sqrt{0.4}$, $m_1 = 0.0030\ eV$ and $\tan \theta_{12} = 0.140$, one gets :

\beq
\label{5.23e}
 \qquad m_2 = -0.0095\ e^{0.0036i}\ eV \qquad m_3 = -0.0495\ e^{0.0075i}\ eV
\eeq
that correspond to the Majorana phases

\beq
\label{5.23bise}
\alpha = {\pi \over 2} - 0.0018 \qquad \qquad \beta = {\pi \over 2} - 0.0038
\eeq

\par \vskip 0.3 truecm

Let us notice an important point. We obtain the CP violating part of the Majorana phases for the light neutrinos (i.e. their departure relatively to ${\pi \over 2}$ in (\ref{5.23bise})) from the $\delta_L$ phase, that we take close to the KM phase $\delta_{KM}$. This can seem paradoxical, because $\delta_L$ concerns the Dirac neutrino mass. However, because of $\delta_L$, the matrix $A^L$ is complex. This implies that, setting $m_1$ real as we have done above, the solutions from eqns. (\ref{5.1e}) for $m_2$ and $m_3$ (with the notation (\ref{2.6-2e})) must be complex. The departure of these Majorana phases relatively to the ones obtained in the real Cabibbo limit (\ref{5.12bise})(\ref{5.16bise}) turn out to be numerically small. \par

We obtain, from (\ref{1.13bise})(\ref{1.13e}) and (\ref{5.23e}) :

\beq
\label{5.23tere}
\Delta m_s^2 = 8.1 \times 10^{-5}\ eV^2 \qquad \qquad \qquad \Delta m_a^2 = 2.4 \times 10^{-3}\ eV^2
\eeq
The agreement with the data is good.\par

Let us now give the complex {\it symmetric} Dirac neutrino mass and the right-handed heavy neutrino spectrum for this solution. We get

\beq
\label{5.24e}
m_D = \left(
        \begin{array}{ccc}
        0.0090+0.0003i & -0.0576-0.0011i & 0.1849+0.0739i \\
        -0.0576-0.0011i & 0.4155-0.0003i & -1.5206+0.0103i \\
        0.1849+0.0739i & -1.5206+0.0103i & 99.9764 \\
        \end{array}
        \right)\ GeV
\eeq
and the quasi-degenerate right-handed heavy neutrino spectrum :
\beq
\label{5.25e}
M_1 = 5.53144 \times 10^9\ GeV \qquad M_2 = 1.43230 \times 10^{10}\ GeV \qquad M_3 = 1.43232 \times 10^{10}\ GeV
\eeq

(2) For $\tan \theta_s \simeq + \sqrt{0.4}$, $m_1 = 0.0062\ eV$ and $\tan \theta_{12} = 0.243$ one gets :

\beq
\label{5.26e}
 \qquad \ \ m_2 = -0.0106\ e^{-0.016i}\ eV \qquad \ \ m_3 = 0.0455\ e^{0.0078i}\ eV
\eeq
that correspond to the Majorana phases

\beq
\label{5.26bis-2e}
\alpha = - {\pi \over 2} + 0.0080 \qquad \qquad \beta = - 0.0039
\eeq

\par \vskip 0.3 truecm

We obtain, from (\ref{1.13bise})(\ref{1.13e}) and (\ref{5.26e}) :

\beq
\label{5.26bise}
\Delta m_s^2 = 7.4 \times 10^{-5}\ eV^2 \qquad \qquad \qquad \Delta m_a^2 = 2.0 \times 10^{-3}\ eV^2
\eeq
The agreement with the data is not as good as for solution (1). We could change the initial conditions for $m_1$ and $\tan \theta_{12}$ and get a better agreement. However, it is not our intention to make a fit but to get a qualitative agreement with the data.\par

We get the Dirac matrix for this solution

\beq
\label{5.27e}
m_D = \left(
        \begin{array}{ccc}
        0.0304+0.0066i & -0.1319-0.0148i & 0.9152+0.3575i \\
        -0.1319-0.0148i & 0.5676-0.0076i & -4.3450+0.0869i \\
        0.9152+0.3575i & -4.3450+0.0869i & 99.8003 \\
        \end{array}
        \right)\ GeV
\eeq
and the quasi-degenerate right-handed heavy neutrino spectrum :
\beq
\label{5.28e}
M_1 = 6.84678 \times 10^9\ GeV \qquad M_2 = 8.84878 \times 10^9\ GeV \qquad M_3 = 8.84909 \times 10^9\ GeV
\eeq

\noindent The signs and phases of the results (\ref{5.23e}) and (\ref{5.26e}) will have quantitative consequences for the effective neutrino mass in neutrinoless double beta decay, as we will see below.

\par \vskip 0.5 truecm

\subsection{CP violation and baryon asymmetry}

\par \vskip 0.5 truecm

The results of the preceding Subsection show that imposing the very drastic conditions (\ref{5.1e}) one gets quasi-degenerate right-handed neutrino spectra.\par 

To have a feeling on how to proceed, let us make an exercise in the case (1), where the quasi-degeneracy (\ref{5.25e}) is less pronounced. Let us compute the Dirac matrix (\ref{4.2e}) in the basis in which the heavy right-handed neutrino mass matrix is diagonal, $\epsilon_1$, ${\tilde m}_1$ and finally $Y_{B_1}$. We will assume that the lightest neutrino decays out-of-equilibrium and that one can apply the one-flavor approximation.

\par \vskip 0.3 truecm

We find the following result

\beq
\label{5.29e}
\hat{m}_D \simeq  \left(
        \begin{array}{ccc}
        -0.055i & -0.052 + 0.132i & -0.131 -0.052i \\
       -0.001 + 0.391i & -0.006 - 1.081i & 1.080 - 0.006i \\
       -0.002 + 0.347i & -0.064 + 70.702i & -70.702 -0.064i \\
        \end{array}
        \right)\ GeV
\eeq
Of course, unlike expression (\ref{5.24e}), this matrix is no longer symmetric. Using it in eqns. (\ref{4.3e}), (\ref{4.6e}) and (\ref{4.8e}), we obtain
\beq
\label{5.30e}
\epsilon_1 \simeq -3.805 \times 10^{-10} \qquad \qquad {\tilde m}_1 \simeq 0.050\ eV \qquad \qquad Y_{B_1} \simeq 3.05 \times 10^{-15} 
\eeq
where we have used the exact formula (\ref{4.4e}) for the function $f(x)$ since the three heavy neutrinos are rather close in mass, although $N_{R_1}$ is lighter, and formula (\ref{4.8e}) is applied because, according to the value of ${\tilde m}_1$ (\ref{5.30e}), we are in the strong washout regime (\ref{4.7e}).\par
 
The obtained baryon asymmetry $Y_{B_1} \cong 3 \times 10^{-15}$ is much too small, by about four to five orders of magnitude, although of the right sign. The reason is the smallnes of $\epsilon_1$, that follows from the quasi-degeneracy of $M_2$ and $M_3$ and the opposite $CP$ asymmetry contribution from both heavy neutrinos. Indeed one finds, for the two terms in (\ref{4.3e}) :

\beq
\label{5.31e}
f\left({M_2^2 \over M_1^2}\right) {\textrm{Im}\left[(\hat{m}_D\hat{m}_D^+)_{12}^2\right] \over 8\pi v^2\ (\hat{m}_D\hat{m}_D^+)_{11}} \simeq -  f\left({M_3^2 \over M_1^2}\right) {\textrm{Im}\left[(\hat{m}_D\hat{m}_D^+)_{13}^2\right] \over 8\pi v^2\ (\hat{m}_D\hat{m}_D^+)_{11}} \simeq - 3.634 \times 10^{-6}
\eeq

\vskip 0.3 truecm

\noindent that shows a strong cancellation giving a very small $CP$ violation $\epsilon_1$.\par

Although for the moment we do not get good phenomenological results, we should however emphasize an interesting limit of the present scheme, namely :
$$\delta_L \to 0 \qquad \textrm{implies} \qquad \epsilon_1 \to 0 \qquad Y_{B_1} \to 0$$
\beq
\label{5.32e}
\alpha = \beta \to {\pi \over 2} \qquad (\textrm{solution (1)}) \qquad \alpha \to - {\pi \over 2}\ , \ \ \beta \to 0 \qquad (\textrm{solution (2)})
\eeq

\vskip 0.3 truecm

Notice that we have an intuitive argument to understand the quasi-degeneracy between $M_2$ and $M_3$ (\ref{5.25e}) and the smallness of the CP violation (\ref{5.30e}). Indeed, in the limit
$A^L_{23} = A^L_{33} = 0$ (\ref{5.1e}), and neglecting terms of order $m_{D_1}^2$ and $m_{D_1}m_{D_2}$, since we take $V^L$ to be close to a diagonal matrix, the $M_R$ matrix has only $M_{R_{13}}$, $M_{R_{31}}$ and $M_{R_{22}}$ sizeable matrix elements with $M_{R_{13}} \simeq M_{R_{31}}$, and the matrix $M_R$ is close to real. Therefore one has $M_2 \simeq M_3$ and $\epsilon_1 \simeq 0$.\par

We can try to modify our very simplified scheme by lifting the degeneracy of $N_{R_2}$ and $N_{R_3}$. We will thus relax somewhat the strong condition (\ref{5.1e}), but keeping the general physical idea of a compact heavy $N_R$ spectrum. We will allow for non-vanishing values for the matrix elements $|A^L_{23}|$ and $|A^L_{33}|$, keeping them "small", i.e. values much smaller than each of their individual contributions that, due to the smallness of the light neutrino masses, are naturally of the order $\sim 10^{11}\ GeV^{-1}$, as can be seen in eqns. (\ref{2.8e})(\ref{2.9e}).\par
As we will further examine, one can thus obtain a rather compact $N_R$ spectrum, and also reasonable values for the baryon asymmetry consistent with the data without spoiling the good properties of the light neutrino spectrum.

\section{Lifting the quasi-degeneracy of heavy neutrinos} \hspace*{\parindent} 

We will proceed now, in terms of some parameters, to a continuous and slow lifting of the quasi-degeneracy of the heavy right-handed neutrino masses obtained in the previous Section within the strong hypothesis (\ref{5.1e}).

In this Section we do the calculation considering non-vanishing values for the r.h.s. of the eqns. (\ref{5.1e}). Moreover, since we have seen that even the drastic assumption of taking $A^L_{33} = A^L_{23} = 0$ gives reasonable neutrino spectra (\ref{5.23e}) or (\ref{5.26e}), we will allow to vary $|A^L_{23}|$ and $|A^L_{33}|$ within a very wide range, keeping "small" values ($<< 10^{11}\ GeV^{-1}$), and observe how the heavy neutrino and the light neutrino spectra evolve, as well as the consequences for the $CP$ violation asymmetry $\epsilon_1$, the effective neutrino mass $\hat{m}_1$ and baryon asymmetry $Y_{B_1}$. We will perform the calculations for both solutions (1) ($\tan \theta_s \simeq - \sqrt{0.4}$) and (2) ($\tan \theta_s \simeq + \sqrt{0.4}$).\par

Notice that the strong conditions (\ref{5.1e}) $A^L_{33} = A^L_{23} = 0$ are linear homogeneous equations in ${1 \over m_i}\ (i = 1,2,3)$. We now allow for non-vanishing inhomogeneous terms 

\beq
\label{6.1e}
A^L_{23}(m_1,m_2,m_3) =  C^L_{23}
\eeq
\beq
\label{6.2e}
A^L_{33}(m_1,m_2,m_3) =  C^L_{33} 
\eeq

To lift the very close degeneracy between $M_2$ and $M_3$ in the case examined before, $A^L_{33} = A^L_{23} = 0$, we just need to have non-vanishing, in general complex parameters $C^L_{23}$ and $C^L_{33}$ in the r.h.s. of (\ref{6.1e}) and (\ref{6.2e}). However, to have an overall compact heavy neutrino spectrum we need inhomogeneous terms that should be small in modulus relatively to each individual term in $A^L_{33}$ and $A^L_{23}$ (\ref{2.8e})-(\ref{2.9e}). This means that we will take non-vanishing values for $C^L_{23}$ and $C^L_{33}$ with the condition 

\beq
\label{6.2bise}
|C^L_{23}|,\  |C^L_{23}| << 10^{11}\ GeV^{-1}
\eeq

In principle one should scan the general two {\it complex} numbers $C^L_{23}$ and $C^L_{33}$ and see how the heavy neutrino spectrum evolves, as well as the light neutrino masses, the light neutrino Majorana phases $\alpha$ and $\beta$, the CP asymmetry $\epsilon_1$ and final baryon asymmetry $Y_{B_1}$.\par 
As pointed out at the beginning of Section 5, the Majorana phases (\ref{5.23bise}) and (\ref{5.26bis-2e}), that we found for $C^L_{23} = C^L_{33} = 0$ have their origin in the approximation adopted for the matrix $V^L$, that we take close to the CKM matrix. In order to preserve this interesting feature, we will assume that the non-vanishing values of the inhomogeneous terms $C^L_{23}$ and $C^L_{33}$ are {\it real} and satisfy (\ref{6.2bise}). Later, in Section 9 we will relax this reality assumption and will see that we have a wide domain of values for complex $C^L_{23}$ and $C^L_{33}$ that can give reasonable results.\par

The first important observation to be made is that the degeneracy between $N_{R_2}$ and $N_{R_3}$, that we have found solving eqns. (\ref{5.1e}) is lifted considerably if $|C^L_{33}|$ is non vanishing in some region with $|C^L_{33}| << 10^{11}\ GeV^{-1}$, and we have realized that the mass difference $M_3 - M_2$ is rather insensitive to the precise value of $|C^L_{23}|$, provided that its value is not "too large". Importantly, the amount of CP violation, and therefore of baryon asymmetry depends also on the value adopted for $|C^L_{23}|$.\par

To reduce the number of parameters, we assume that $C^L_{23}$ and $C^L_{33}$ are real and of equal modulus $|C^L_{23}| = |C^L_{33}|$. As we vary $C^L_{23}$ and $C^L_{33}$ we find essentially the same heavy neutrino spectrum and roughly the same values for $\Delta m_s^2$ and $\Delta m_a^2$, independently of their relative sign. We find that what is dependent on this relative sign is the amount of CP violation and baryon asymmetry. If $C^L_{23} = C^L_{33}$ (independently of its sign), the baryon asymmetry can be at most of $O(10^{-12})$, but if $C^L_{23}$ and $C^L_{33}$ are of opposite sign, one can get a correct amount of baryon asymmetry.

In conclusion, after some trial and error guesses, we respectively adopt real numbers for $C^L_{23}$ and $C^L_{33}$ for both solutions (1) ($\tan \theta_s < 0$) and (2) ($\tan \theta_s > 0$) 

\beq
\label{6.3e}
(1) \qquad \qquad \qquad \qquad - C^L_{23} = C^L_{33} > 0 \qquad \qquad  \qquad \qquad 
\eeq
\beq
\label{6.4e}
(2) \qquad \qquad \qquad \qquad - C^L_{23} = C^L_{33} < 0 \qquad \qquad  \qquad \qquad 
\eeq
with $|C^L_{33}| << 10^{11}\ GeV$. As it will become clear below, the adopted sign for each of the solutions corresponds to the experimental sign $Y_B > 0$ in some region for the parameters $C^L_{23}, C^L_{33}$. We assume that $Y_{B_1} \simeq Y_B$, an hypothesis that will be justified in Section 10.\par 
We will now show how the heavy neutrino and the light neutrino spectra evolve under the conditions (\ref{6.3e})(\ref{6.4e}). In the next Section we will show how $\epsilon_1$, $\hat{m}_1$ and $Y_{B_1}$ behave.\par
Of course, our ansatz for $C^L_{23}$, $C^L_{33}$ is just a guess. We do not intend to make a fit to the overall data, light neutrino spectrum and baryon asymmetry. We just want to see if the description of these data is possible within this scheme of a compact heavy neutrino spectrum and approximate quark-lepton symmetry. Other equations of the type (\ref{6.1e})(\ref{6.2e}), with {\it complex} r.h.s. values for the parameters $C^L_{23}$,  $C^L_{33}$ give also acceptable results, as we will see in Section 9. 

\vskip 1.0 truecm

\subsection{Heavy and light neutrino spectra for case (1) $\tan \theta_s < 0$}

\vskip 0.5 truecm

To perform the calculations we adopt again the values of the pure Cabibbo limit, namely $m_1 = 0.0030\ eV$, $\tan \theta_{12} = 0.140$, although one could slightly change these values to get a better fit.

\vskip 0.5 truecm

In Fig. 1-a the right-handed heavy neutrino spectrum is plotted. In Figures 1-b and 1-c we show respectively the solar and atmospheric quantities $\Delta m_s^2$, $\Delta m_a^2$.
\par \vskip 0.5 truecm

\includegraphics[scale=1.5]{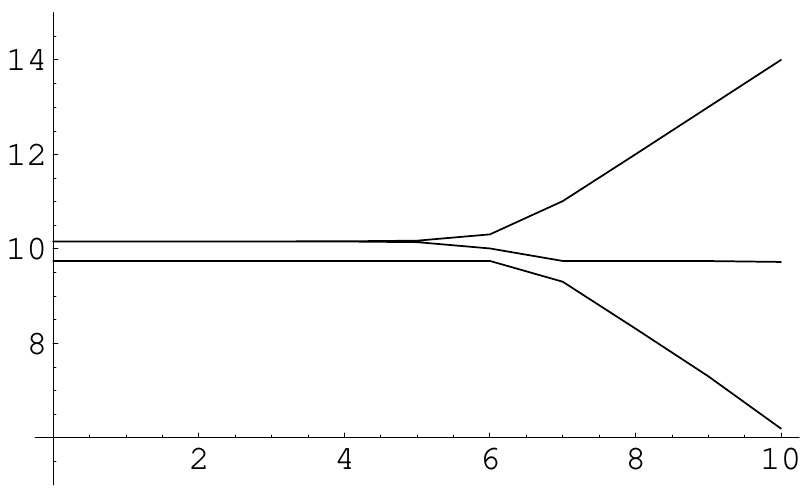}

\vskip 0.5 truecm

Fig. 1-a. Log-log plot of the right-handed heavy neutrino spectrum (masses in GeV units) as a function of $- C^L_{23} = C^L_{33} > 0$ in units of $GeV^{-1}$, for fixed $m_1 = 0.0030\ eV$ and $\tan \theta_{12} =  0.140$. Within the range $- C^L_{23} = C^L_{33} = 10^6 - 10^7\ GeV^{-1}$ there is a level crossing. The angular points come from the fact that the curves are obtained from interpolation of a finite number of points. The same applies to the other figures.

\par \vskip 1.0 truecm

\includegraphics[scale=1.5]{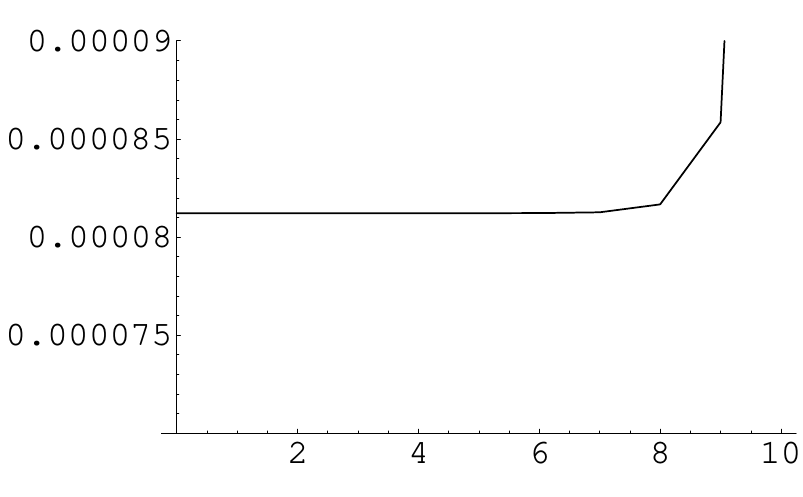}

Fig. 1-b. $\Delta m_s^2$ in $eV^2$ units as a function of $- C^L_{23} = C^L_{33} > 0$ in units of $GeV^{-1}$, for fixed $m_1 = 0.0030\ eV$ and $\tan \theta_{12} =  0.140$, in a log scale for $C^L_{33}$.

\par \vskip 1.0 truecm

\includegraphics[scale=1.5]{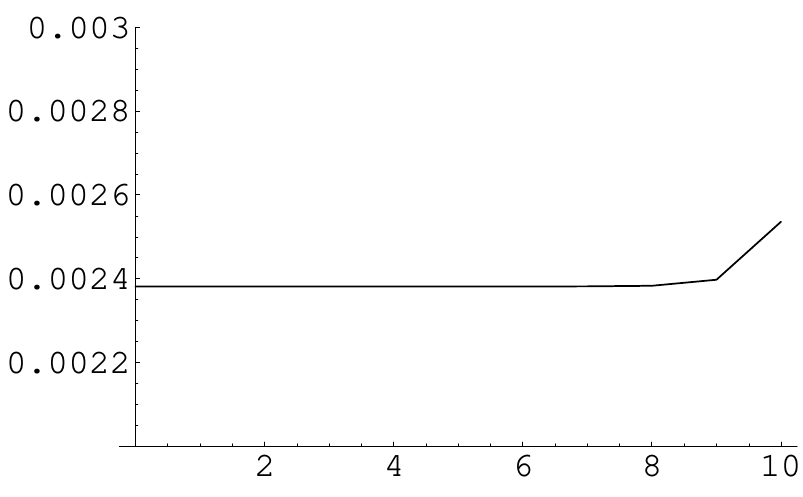}

Fig. 1-c. $\Delta m_a^2$ in $eV^2$ units as a function of $- C^L_{23} = C^L_{33} > 0$ in units of $GeV^{-1}$, for fixed $m_1 = 0.0030\ eV$ and $\tan \theta_{12} =  0.140$, in a log scale for $C^L_{33}$.

\par \vskip 1.0 truecm

Let us comment on these figures. The angular points that appear in the figures are an artifact of the representation of the curves, obtained from an interpolation of a finite number of points. Notice that {\it for each point} we must perform the singular value decomposition of the matrix $M_R$ in order to compute the quantities necessary to obtain the baryon asymmetry.\par 
The first striking point is that, as we have learned from eqns. (\ref{6.1e})(\ref{6.2e}), the $N_R$ spectrum (Fig. 1-a) is very compact for $C^L_{33}$ not "too large", $C^L_{33} < 10^7\ GeV^{-1}$. As explained in the Introduction, there is an expected correlation between the stability of the light neutrino spectrum and the compact heavy right-handed neutrino one. The fine-tuning for the close heavy neutrino masses ensures the stability of the light neutrino ones. For $- C^L_{23} = C^L_{33} > 10^7\ GeV^{-1}$ the right-handed neutrino spectrum evolves into a hierarchical spectrum. The values obtained for $\Delta m_s^2$ (Fig. 1-b) and $\Delta m_a^2$ (Fig. 1-c) are very stable and consistent with experiment for a wide range of values of $- C^L_{23} = C^L_{33}$, of about eight orders of magnitude.\par
We observe two other important things in Fig. 1-a : the degeneracy between $N_{R_2}$ and $N_{R_3}$ is lifted, and there is a level crossing around $- C^L_{23} \simeq C^L_{33} = 3 \times 10^6\ GeV^{-1}$.\par
An important point to be also underlined is that one of the levels ($N_{R_1}$ before the crossing) remains practically constant in the whole studied range, while the mass of $N_{R_2}$ decreases. After the crossing we will call $N_{R_1}$ this right-handed neutrino, being the lightest, according to convention (\ref{4.2bise}).\par

As we can see from Figs. 1-b and 1-c, the values obtained for $\Delta m_s^2$ and $\Delta m_a^2$ are in good agreement with the data for a very wide range of the parameters.

\par \vskip 1.0 truecm

For the Majorana phases $\alpha$ and $\beta$, shown in Figs. 1-d and 1-e we find rather constant values (in a logarithmic scale) that are very close but a little smaller than ${\pi \over 2}$, as shown in the figures below.

\par \vskip 1.0 truecm

\includegraphics[scale=1.5]{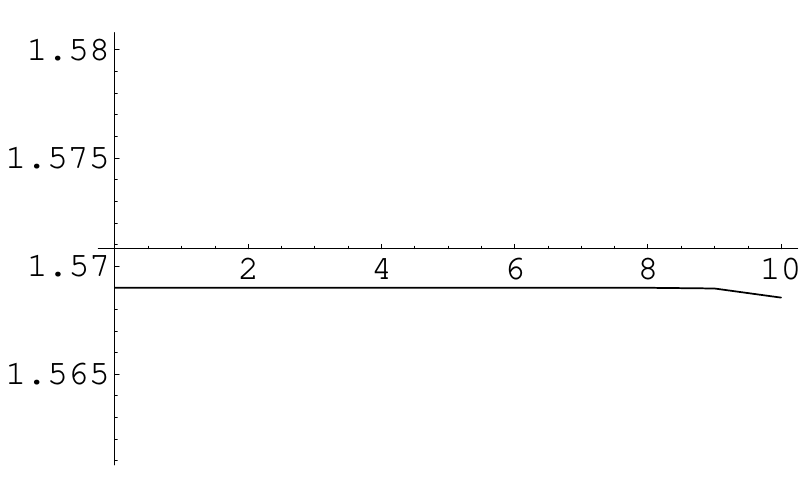}

Fig. 1-d. The Majorana phase $\alpha$ as a function of $- C^L_{23} = C^L_{33} > 0$ in units of $GeV^{-1}$, in a log scale for $C^L_{33}$, for fixed $m_1 = 0.0030\ eV$ and $\tan \theta_{12} =  0.140$. The x-axis is centered at $\pi/2$.

\par \vskip 1.0 truecm

\includegraphics[scale=1.5]{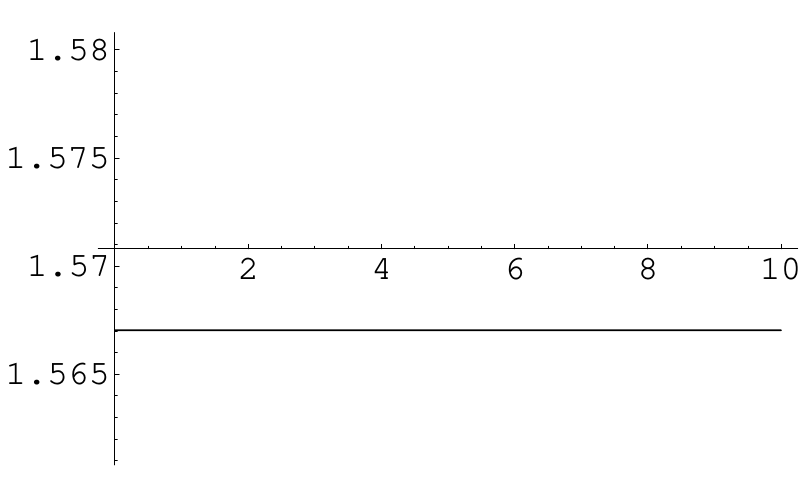}

Fig. 1-e. The Majorana phase $\beta$ as a function of $- C^L_{23} = C^L_{33} > 0$ in units of $GeV^{-1}$, in a log scale for $C^L_{33}$, for fixed $m_1 = 0.0030\ eV$ and $\tan \theta_{12} =  0.140$. The x-axis is centered at $\pi/2$.

\par \vskip 1.0 truecm

\subsection{Heavy and light neutrino spectra for case (2) $\tan \theta_s > 0$}

\vskip 0.5 truecm 

To perform the calculations we adopt again the values obtained in the pure Cabibbo limit, namely $m_1 = 0.0062\ eV$, $\tan \theta_{12} = 0.243$.

In Fig. 2-a the right-handed heavy neutrino spectrum is plotted. In Figures 2-b and 2-c we show respectively the solar and atmospheric quantities $\Delta m_s^2$, $\Delta m_a^2$. Figs. 2-d and 2-e display the result for the Majorana phases $\alpha$ and $\beta$.

\par \vskip 1.0 truecm 

\includegraphics[scale=1.5]{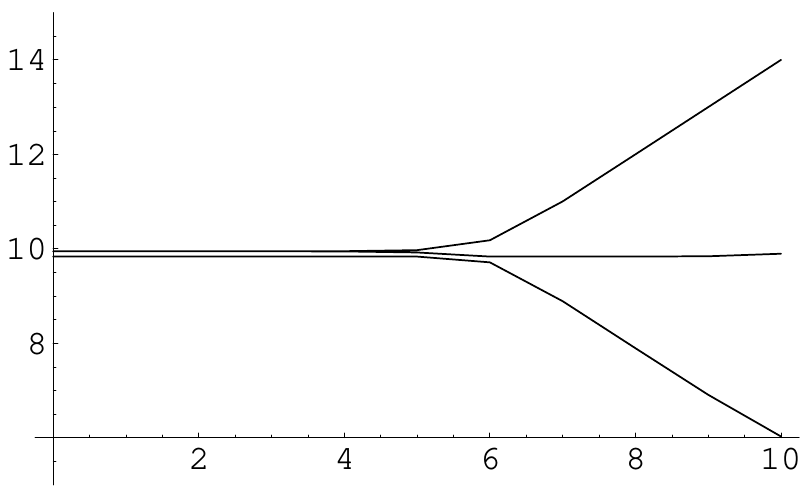}

Fig. 2-a. Log-log plot of the right-handed heavy neutrino spectrum (masses in GeV units) as a function of $C^L_{23} = - C^L_{33} > 0$ in units of $GeV^{-1}$, for fixed $m_1 = 0.0062\ eV$ and $\tan \theta_{12} =  0.243$. Within the range $- C^L_{33} = 10^5 - 10^6\ GeV^{-1}$ there is a level crossing.

\par \vskip 1.0 truecm

\includegraphics[scale=1.5]{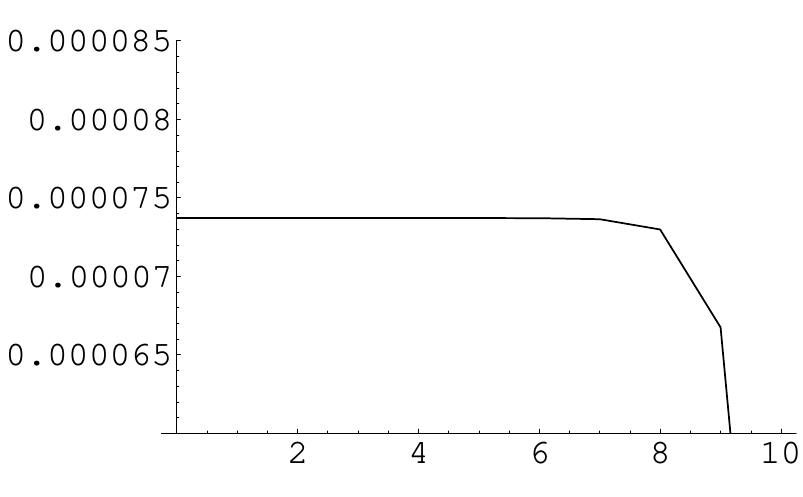}

Fig. 2-b. $\Delta m_s^2$ in $eV^2$ units as a function of $C^L_{23} = - C^L_{33} > 0$ in units of $GeV^{-1}$, in a log scale for $- C^L_{33}$, for fixed $m_1 = 0.0062\ eV$ and $\tan \theta_{12} =  0.243$.

\par \vskip 0.5 truecm

\includegraphics[scale=1.5]{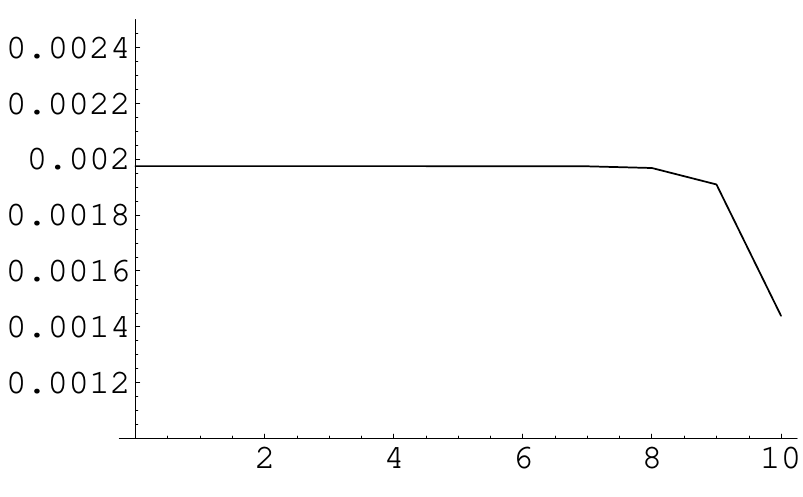}

Fig. 2-c. $\Delta m_a^2$ in $eV^2$ units as a function of $C^L_{23} = - C^L_{33} > 0$ in units of $GeV^{-1}$, in a log scale for $- C^L_{33}$, for fixed $m_1 = 0.0062\ eV$ and $\tan \theta_{12} =  0.243$.

\par \vskip 0.5 truecm

As shown in the figures below, in a logarithmic scale, we find for the Majorana phase $\alpha$ a rather constant value that is very close but a little larger than $- {\pi \over 2}$ and for $\beta$ a small negative almost constant value.

\par \vskip 0.5 truecm

\includegraphics[scale=1.5]{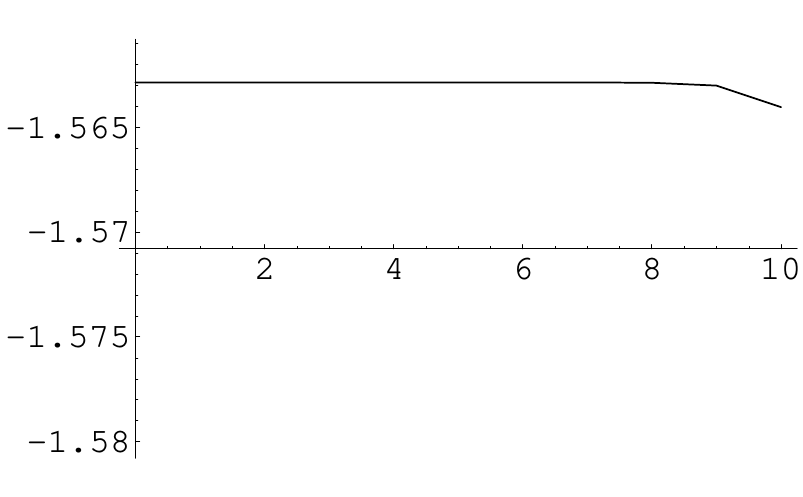}

Fig. 2-d. The Majorana phase $\alpha$ as a function of $C^L_{23} = - C^L_{33} > 0$ in units of $GeV^{-1}$, in a log scale for $- C^L_{33}$, for fixed $m_1 = 0.0062\ eV$ and $\tan \theta_{12} =  0.243$. The x-axis is centered at $- \pi/2$.

\par \vskip 1.0 truecm

\includegraphics[scale=1.5]{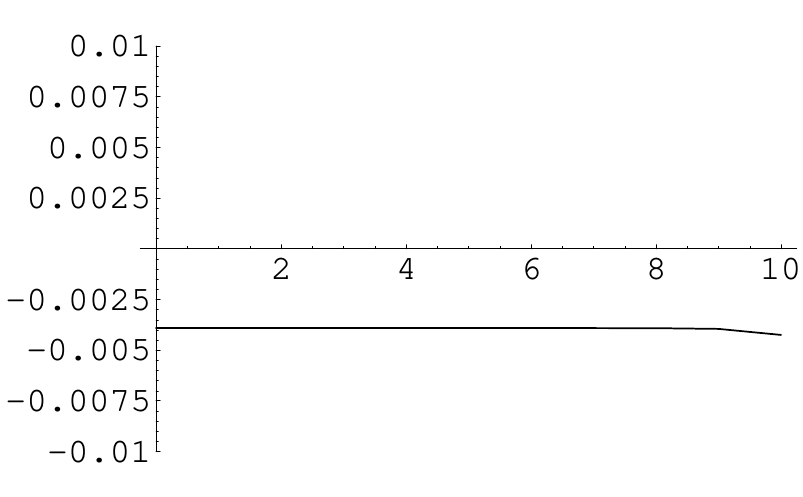}

Fig. 2-e. The Majorana phase $\beta$ as a function of $C^L_{23} = - C^L_{33} > 0$ in units of $GeV^{-1}$, in a log scale for $- C^L_{33}$, for fixed $m_1 = 0.0062\ eV$ and $\tan \theta_{12} =  0.243$.

\par \vskip 1.0 truecm

The first striking point in this case is that, as imposed from eqns. (\ref{6.1e})(\ref{6.2e}), the $N_R$ spectrum (Fig. 2-a) has the same features as for the precedent solution, although it is very compact for $- C^L_{33} < 10^6\ GeV^{-1}$, much more than in case (1). Within the range $C^L_{23} = - C^L_{33} = 10^5 - 10^6\ GeV^{-1}$ there is also a level crossing, on which we will comment below. For $C^L_{23} = - C^L_{33} > 10^6\ GeV^{-1}$ the right-handed neutrino spectrum evolves also into a hierarchical spectrum. Secondly, $\Delta m_s^2$ (Fig. 2-b) and $\Delta m_a^2$ (Fig. 2-c) are very stable for a wide range of values of $C^L_{23} = - C^L_{33}$, of about seven order of magnitude. However, the agreement with experiment is not as good as for solution (1), although it is acceptable within a $3 \sigma$ range. Of course, we could somewhat change the initial conditions $m_1 = 0.0062\ eV$, $\tan \theta_{12} = 0.243$ and become closer to the data. This could be done, but we will not do it because our purpose is only a qualitative one within our (fine-tuning) scheme.\par

\newpage

\section{CP violation and baryon asymmetry in the region approaching the level crossing} \hspace*{\parindent} 

Let us turn now to the quantities that are important for Baryogenesis via Leptogenesis. Labelling the lightest heavy right-handed neutrino $N_{R_1}$, our calculations show that the quantities $- \epsilon_1$, $\tilde{m}_1$ and $Y_{B_1}$ have a strong discontinuity across the level crossing region. We call always $N_{R_1}$ the lightest heavy neutrino, even after the crossing, according to the convention (\ref {4.2bise}). \par

We will justify and characterize this term of level crossing, and discuss its implications before this region in Section 10. \par

To simplify the presentation of the results, we will restrict ourselves to the region {\it before the crossing}, where $M_2$ and $M_1$ become relatively close, i.e., to the following regions, slightly different in both cases :
$$ (1) \qquad \qquad \qquad \qquad 10^5\ GeV^{-1} \leq - C^L_{23} = C^L_{33} \leq 10^{6.4}\ GeV^{-1} \qquad \qquad \qquad \qquad $$  
\beq
\label{7.1e}
10^9\ GeV \leq M_2 - M_1 \leq 8.3 \times 10^9\ GeV
\eeq

$$ (2) \qquad \qquad \qquad \qquad 10^4\ GeV^{-1} \leq C^L_{23} = - C^L_{33}  \leq 10^{5.6}\ GeV^{-1}  \qquad \qquad  \qquad \qquad $$
\beq
\label{7.2e}
10^9\ GeV \leq M_2 - M_1 \leq 2. \times 10^9\ GeV 
\eeq

To avoid the delicate situation related to the quasi-degeneracy of two heavy neutrinos, extensively studied by A. Pilaftsis et al. \cite{PILAFTSIS}, we need the condition (see also \cite{AFS})
\beq
\label{7.3e}
\Gamma_1 << M_2-M_1
\eeq
where $\Gamma_1$ is the width of the lightest heavy right-handed neutrino, that has an upper bound qualitatively given by \cite{AFS} :

\beq
\label{7.4e}
\Gamma_1 \leq {m_t^2 \over 16 \pi v^2} M_1 
\eeq

\noindent Before the level crossing region one gets, from the parameters quoted above ($m_t \simeq m_{D_3} \simeq 100\ GeV$, $v = 174\ GeV$ and $M_1 \simeq  5. \times 10^9\ GeV$) :

\beq
\label{7.5e}
\Gamma_1 \leq 3. \times 10^7\ GeV
\eeq

\noindent Therefore, before the level crossing region, taking into account the inequalities (\ref{7.1e})(\ref{7.2e}), we see that in both cases (1) and (2) the condition (\ref{7.3e}) is satisfied. We are far away from resonant leptogenesis and we do not have to face complications related to quasi-degenerate heavy neutrinos, for which $M_2-M_1 \leq \Gamma_1$.

\par \vskip 0.5 truecm

Let us now show the quantities $- \epsilon_1$, $\tilde{m}_1$ and $Y_{B_1}$ for both solutions within the interesting ranges (\ref{7.1e})(\ref{7.2e}).

\par \vskip 0.5 truecm

\subsection{Case (1) $\tan \theta_s < 0$} 

\par \vskip 0.5 truecm

Fig. 1-d displays the $CP$ violation parameter $- \epsilon_1$, Fig. 1-e the washout parameter $\tilde{m}_1$, and in Fig. 1-f the baryon asymmetry $Y_{B_1}$ in the one-flavor approximation.

\par \vskip 1. truecm

\includegraphics[scale=1.5]{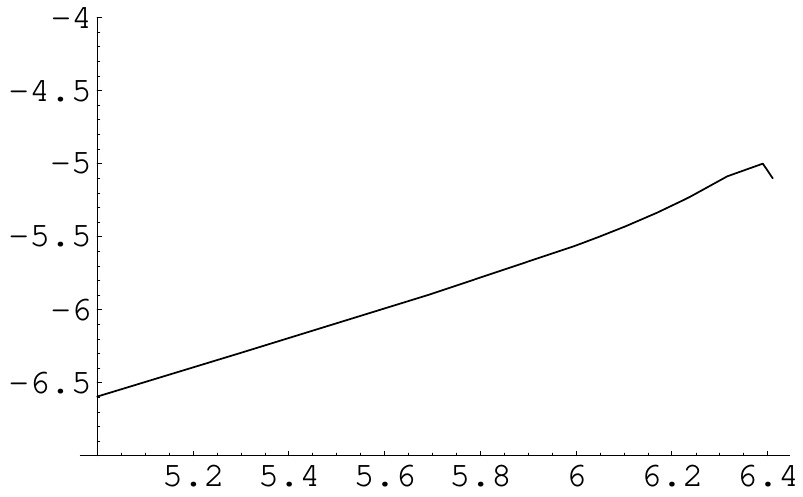}

Fig. 1-d. Log-log plot of $- \epsilon_1$ as a function of $- C^L_{23} = C^L_{33}$ in units of $GeV^{-1}$, for fixed $m_1 = 0.0030\ eV$ and $\tan \theta_{12} =  0.140$.  

\par \vskip 1.0 truecm

\includegraphics[scale=1.5]{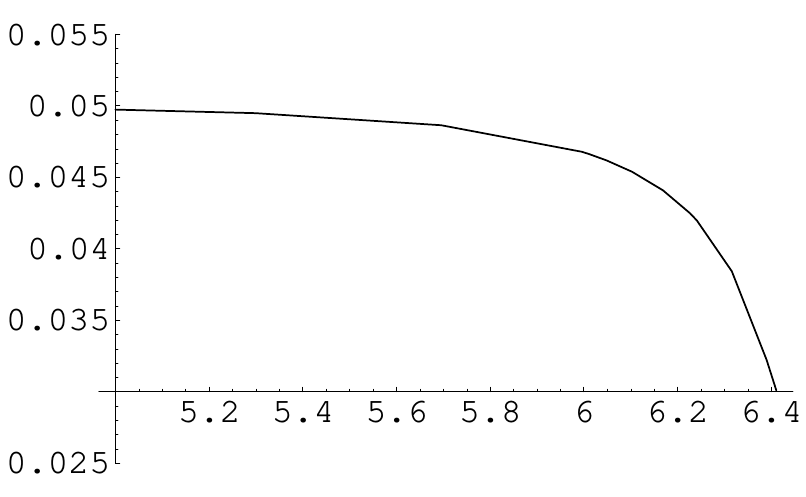}

Fig. 1-e. $\tilde{m}_1$ in $eV$ units as a function of $- C^L_{23} = C^L_{33}$ in units of $GeV^{-1}$, in a log scale for $C^L_{33}$, for fixed $m_1 = 0.0030\ eV$ and $\tan \theta_{12} =  0.140$. 

\par \vskip 1.0 truecm

\includegraphics[scale=1.5]{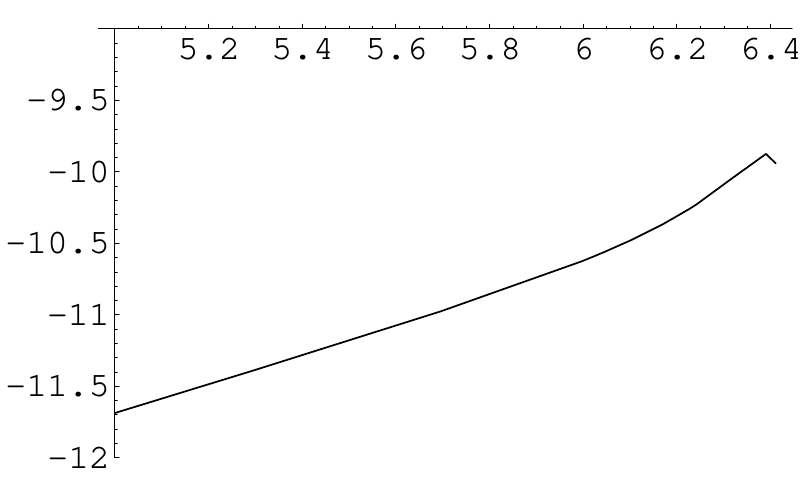}

Fig. 1-f. Log-log plot of $Y_{B_1}$ as a function of $- C^L_{23} = C^L_{33}$ in units of $GeV^{-1}$, for fixed $m_1 = 0.0030\ eV$ and $\tan \theta_{12} =  0.140$.

\par \vskip 1.0 truecm

We observe that $\epsilon_1$ is negative and becomes large enough in absolute magnitude to give a large positive $Y_{B_1}$ with rather stable values of the parameter $\tilde{m}_1$ that imply strong wash-out in the whole region. Notice that $- \epsilon_1$ as well as $Y_{B_1}$ grow as the mass difference $M_3-M_2$ slowly grows and the mass difference $M_2-M_1$ becomes smaller, a phenomenon already underlined by Akhmedov et al. \cite{AFS}.\par
The main conclusion that we can draw from Fig. 1-f is that there is no problem to get a baryon asymmetry $Y_{B_1}$ of the right order of magnitude $(Y_{B})_{exp} \simeq 9 \times 10^{-11}$.\par
We must underline that if we did took the opposite sign in (\ref{6.3e}), we would have obtained the opposite sign for $\epsilon_1$ and therefore also for $Y_{B_1}$. Therefore, our scheme {\it does not predict} the sign of $Y_B$ since it depends on the chosen sign of the inhomogeneous terms.
   
\par \vskip 1.0 truecm

\subsection{Case (2) $\tan \theta_s > 0$} 

\par \vskip 0.5 truecm

Fig. 2-d displays the $CP$ violation parameter $- \epsilon_1$, Fig. 2-e the washout parameter $\tilde{m}_1$, and in Fig. 2-f the baryon asymmetry $Y_{B_1}$ in the one-flavor approximation.

\par \vskip 2.0 truecm

\includegraphics[scale=1.5]{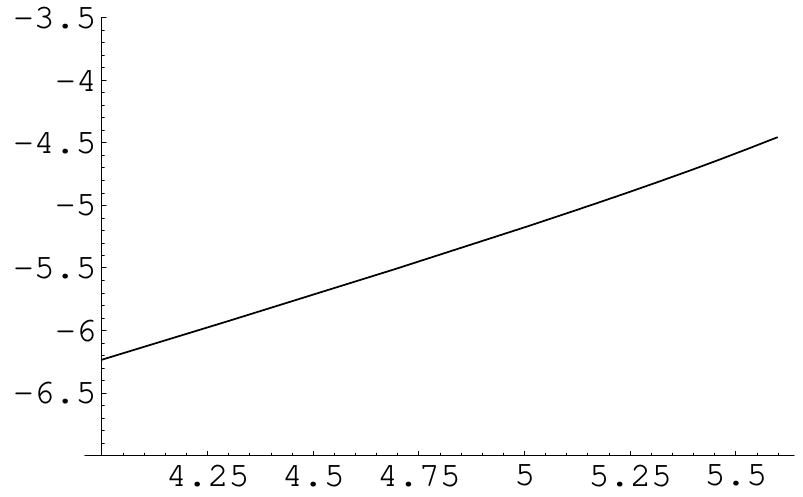}

Fig. 2-d. Log-log plot of $- \epsilon_1$ as a function of $C^L_{23} = - C^L_{33}$ in units of $GeV^{-1}$, for fixed $m_1 = 0.0062\ eV$ and $\tan \theta_{12} =  0.243$.  

\par \vskip 1.0 truecm

\includegraphics[scale=1.5]{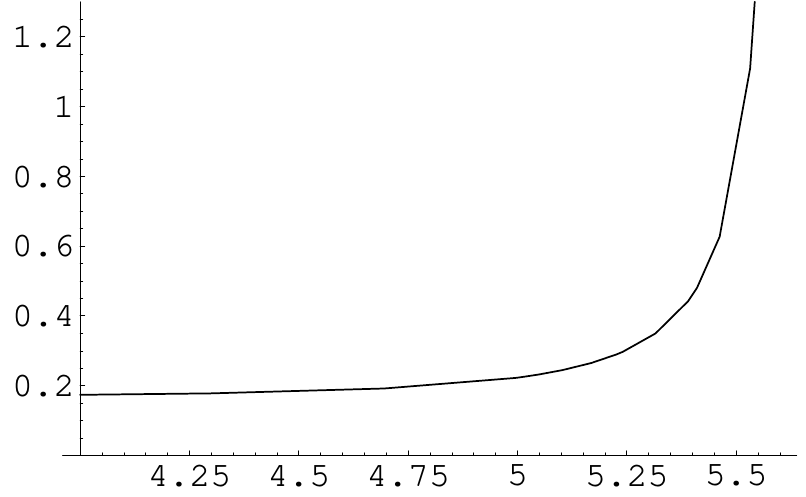}

Fig. 2-e. $\tilde{m}_1$ in $eV$ units as a function of $C^L_{23} = - C^L_{33}$ in units of $GeV^{-1}$, in a log scale for $- C^L_{33}$, for fixed $m_1 = 0.0062\ eV$ and $\tan \theta_{12} =  0.243$. 

\par \vskip 1.0 truecm

\includegraphics[scale=1.5]{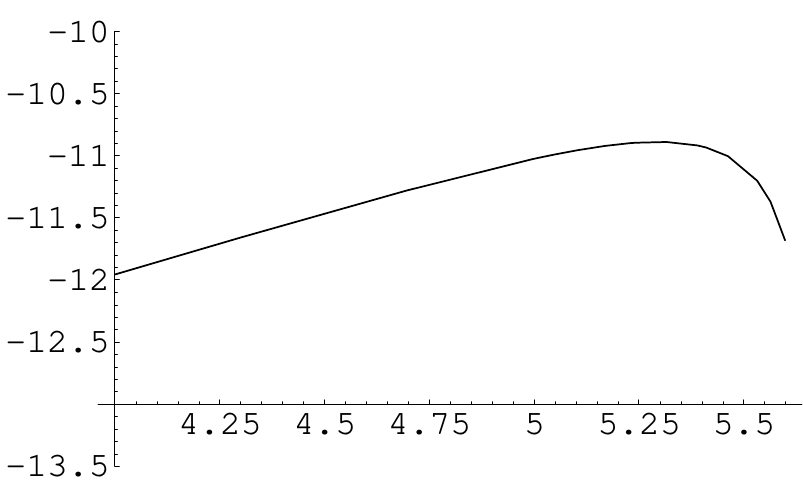}

Fig. 2-f. Log-log plot of $Y_{B_1}$ as a function of $C^L_{23} = - C^L_{33}$ in units of $GeV^{-1}$, for fixed $m_1 = 0.0062\ eV$ and $\tan \theta_{12} =  0.243$.

\par \vskip 1.0 truecm

We observe that $\epsilon_1$ is negative and becomes large in absolute magnitude to give a rather large positive $Y_{B_1}$ with values of the parameter $\tilde{m}_1$ within the strong wash-out regime.\par
We must again point out that if we did took the opposite signe of the r.h.s. of (\ref{6.4e}), we would have obtained the opposite sign for $\epsilon_1$ and therefore also for $Y_{B_1}$.\par
The main conclusion that we can draw from Fig. 2-f is that in this case we are somewhat short of having a baryon asymmetry $Y_{B_1}$ of the right order of magnitude $(Y_{B})_{exp} \simeq 9 \times 10^{-11}$. However, as pointed out above, one could modify the initial conditions (the values $m_1 = 0.0062$ and $\tan \theta_{12} = 0.243$) and get results in better agreement with the data. But it is not our purpose to make a detailed fit for $\Delta m_s^2$, $\Delta m_a^2$ and $Y_{B_1}$, we want just to give a qualitative trend.
\par \vskip 0.3 truecm

Let us emphasize again that in the present scheme developped in Sections 6 and 7, due to the reality conditions on $C^L_{23}$ and $C^L_{33}$ we have again the interesting limit (\ref{5.32e}) :
$$\delta_L \to 0 \qquad \textrm{implies} \qquad \epsilon_1 \to 0 \qquad Y_{B_1} \to 0$$
\beq
\label{7.10e}
\alpha = \beta \to {\pi \over 2} \qquad (\textrm{solution (1)}) \qquad \alpha \to - {\pi \over 2}\ , \ \ \beta \to 0 \qquad (\textrm{solution (2)})
\eeq

\section{Results for $m_{\nu_e}$, $(\beta \beta)_{0\nu}$ and sum of neutrino masses} \hspace*{\parindent} 

We give here the predictions for the electron neutrino mass, on which one has limits from tritium $\beta$ decay :
\beq
\label{8.1e}
m_{\nu_e} = \cos^2\theta_s\ |m_1| + \sin^2\theta_s\ |m_2|
\eeq
and for the effective mass relevant for neutrinoless double beta decay $|< m_{ee} >|$ that writes, within the approximation (\ref{2.6-1e}),
\beq
\label{8.2e}
|< m_{ee} >|\ = |\cos^2\theta_s\ m_1 + \sin^2\theta_s\ m_2|
\eeq

The neutrino masses and their phases for both solutions (\ref{5.23e})(\ref{5.26e}) are very close to those obtained in the region that give an acceptable value for $Y_B$. Taking thus  the values for both solutions, using the notation (\ref{2.6-2e}) :
\beq
\label{8.3e}
(1) \ \ m_1 = 0.0030\ eV, \ \ \ m_2 = - 0.0095\ e^{0.0036i}\ eV, \ \ \ m_3 = - 0.0495\ e^{0.0075i}\ eV
\eeq
\beq
\label{8.4e}
(2) \ \ m_1 = 0.0062\ eV, \ \ \ m_2 = - 0.0106\ e^{-0.016i}\ eV, \ \ \ m_3 = 0.0455\ e^{0.0078i}\ eV
\eeq
we obtain, respectively :
\beq
\label{8.5e}
(1) \qquad \qquad m_{\nu_e} \simeq 4.9 \times 10^{-3}\ eV \qquad \qquad |< m_{ee} >|\ \simeq   5.7 \times 10^{-4}\ eV\qquad
\eeq
\beq
\label{8.6e}
(2) \qquad \qquad m_{\nu_e} \simeq 7.5 \times 10^{-3}\ eV \qquad \qquad |< m_{ee} >|\ \simeq  1.4 \times 10^{-3}\ eV\qquad
\eeq
For both solutions, due to the relative signs among the $m_i$ (i = 1, 2, 3) (i.e., due to the Majorana phases), there is a strong cancellation between the two terms in (\ref{8.2e}), a phenomenon already exhibited in \cite{BF} in another context. The cancellation is stronger for solution (1).\par
For the sum of the absolute magnitude of all neutrino masses, we obtain :

\beq
\label{8.7e}
(1) \qquad \qquad \qquad \sum_i |m_{\nu_i}| = 0.0620\ eV \qquad \qquad
\eeq
\beq
\label{8.8e}
(2) \qquad \qquad \qquad \sum_i |m_{\nu_i}| = 0.0623\ eV \qquad \qquad
\eeq

\noindent One gets very close results for both solutions, that comply with the cosmological bounds (\ref{1.8e}) and (\ref{1.8bise}) \cite{FLMMPSSS}.\par

Let us make a last qualitative remark comparing the different possible future experiments on neutrino masses and stress the importance of cosmological limits.\par 
If one takes $m_1 \simeq 0$ one finds, from the data, $|m_2| \simeq \sqrt{\Delta m_s^2} \simeq 9 \times 10^{-3}\ eV$ and $|m_3| \simeq \sqrt{\Delta m_a^2+\textrm{cos}^2\theta_s\Delta m_s^2} \simeq 5 \times 10^{-2}\ eV$, which correspond to a value for the l.h.s. in eqns. (\ref{1.8e})(\ref{1.8bise}) of the order $6 \times 10^{-2}\ eV$, very near the value that we have found and only a factor 3.3 below the bound (\ref{1.8bise}) \cite{FLMMPSSS}. So, according to the present scenario, the most promising search for effects of neutrino masses, apart from oscillation experiments, is the analysis of cosmological data, while for beta decay and neutrinoless double beta decay one should need an improvement of more than two orders of magnitude.

\section{Relaxing the additional reality constraints of the model}
\hspace*{\parindent}

We now relax the conditions of the particular model that we have studied quantitatively, namely given by eqs. (\ref{6.1e})(\ref{6.2e}) with real $C^L_{23}$ and $C^L_{33}$ satisfying (\ref{6.3e})(\ref{6.4e}), and allow complex numbers for these parameters, keeping however "small" values for the moduli, as stated in (\ref{6.2bise}).\par
Notice the important point that now we have  two new sources of CP violation besides a single independent phase $\delta_L$ (or $\alpha$ or $\beta$), and we recover the situation in which there are three independent phases : $\delta_L$, $\alpha$ and $\beta$. However, as we will see below, in the range of interest for $Y_{B_1}$, the magnitude of the Majorana phases is not very different than in the real case studied in detail in Sections 5, 6 and 7. However, their new contributions have important implications for the baryon asymmetry.\par
Just to have a feeling of what can happen, we take an extreme case and adopt relations (\ref{6.3e})(\ref{6.4e}) with the condition (\ref{6.2bise}), but taking now $C^L_{23}$ and $C^L_{33}$ {\it purely imaginary}. Interestingly, the results are phenomenologically good and show that our general scheme of a compact $N_R$ spectrum is flexible enough.\par
We do not give the corresponding curves of Sections 6 and 7, and give values for representative points with acceptable phenomenological results.\par
For case (1), i.e. $\tan \theta_s \simeq - \sqrt{0.4}$ with $m_1 = 0.0030\ eV$ and $\tan \theta_{12} =  0.140$, taking

\beq
\label{9.1e}
- C^L_{23} = C^L_{33} = i\ 10^{-5}
\eeq

\noindent we find the following results :
$$\Delta m_s^2 = 8.1 \times 10^{-5}\ eV^2 \qquad \qquad \qquad \Delta m_a^2 = 2.4 \times 10^{-3}\ eV^2$$
$$\alpha = {\pi \over 2} - 0.0018 \qquad \qquad \qquad \qquad \qquad \beta = {\pi \over 2} - 0.0038$$
$$M_1 = 5.531 \times 10^9\ GeV \qquad M_2 = 1.383 \times 10^{10}\ GeV \qquad M_3 = 1.483 \times 10^{10}\ GeV$$
\beq
\label{9.2e}
\tilde{m}_1 = 0.050\ eV \qquad \qquad \epsilon_1 = - 2.755 \times 10^{-5} \qquad \qquad Y_{B_1} = 2.211 \times 10^{-10}
\eeq

\vskip 1.0 truecm

\noindent while for case (2), i.e.  $\tan \theta_s \simeq + \sqrt{0.4}$ with $m_1 = 0.0062\ eV$ and $\tan \theta_{12} =  0.243$, taking
\beq
\label{9.3e}
- C^L_{23} = C^L_{33} = -i\ 10^{-3}
\eeq

\noindent we find :
$$\Delta m_s^2 = 7.4 \times 10^{-5}\ eV^2 \qquad \qquad \qquad \Delta m_a^2 = 2.0 \times 10^{-3}\ eV^2$$
$$\alpha = - {\pi \over 2} + 0.0079 \qquad \qquad \qquad \qquad \qquad \beta = - 0.0039$$
$$M_1 = 6.847 \times 10^9\ GeV \qquad M_2 = 8.844 \times 10^{9}\ GeV \qquad M_3 = 8.954 \times 10^{9}\ GeV$$
\beq
\label{9.4e}
\tilde{m}_1 = 0.170\ eV \qquad \qquad \epsilon_1 = - 3.622 \times 10^{-5} \qquad \qquad Y_{B_1} = 7.035 \times 10^{-11}
\eeq

These results are phenomenologically reasonable, and we find a whole region in their neighborhood that gives also good results.\par Notice that in the numbers of case (2) we are not far away from saturating the bound (\ref{7.5e}), and therefore we are approaching the regime of resonant leptogenesis.\par
Let us emphasize again that in the case examined here we have two different sources of CP violation : $\delta_L$ and the Majorana phases $\alpha$, $\beta$.\par 
To illustrate how these new contributions to Majorana phases occur, it is useful to recall again how we perform our calculations. Proceeding like in Section 6, using eqns. (\ref{9.1e}) and (\ref{9.3e}), we compute, from (\ref{2.8e})(\ref{2.9e}), $m_2$ and $m_3$ (with the convention (\ref{2.6-2e})) in terms of the given values for $m_1$ and $\tan \theta_{12}$. Then, $m_2$ and $m_3$ get by construction new CP-violating contributions to the Majorana phases because, according to (\ref{9.1e})(\ref{9.3e}), the inhomogeneous terms must be pure imaginary.\par 
Of course, since now we have new sources of CP violation in the Majorana phases, in the limit $\delta_L \to 0$ we do not recover the simple limit (\ref{7.10e}) that we got for real values of $C^L_{23}$ and $C^L_{33}$ or, equivalently, for CP violation in the Majorana phases fixed exclusively from the phase $\delta_L$.\par
We had CP violation in Majorana phases that were induced by their calculation for a given $\delta_L$ in the case of vanishing $C^L_{23}$ and $C^L_{33}$ (Subsection 5.2). But, from the imaginary inhomogeneous terms of the present Section, we have now new sources of CP violation in these phases.\par
These new sources of CP violation in the Majorana phases, although small, are very efficient in producing a baryon asymmetry, as we realize from the results (\ref{9.2e}) and (\ref{9.4e}).\par 
It can be easily understood that the constraints (\ref{9.1e}) and (\ref{9.3e}) imply new contributions to the baryon asymmetry. These equations mean that the entries $A^L_{23} = A^L_{32}$ and $A^L_{33}$ are purely imaginary. This in turn implies, from (\ref{3.4e}), new CP violation contributions to the mass matrix $M_R$, providing, after its diagonalization, new contributions to $\epsilon_1$ and $Y_{B_1}$.
This important phenomenon certainly deserves further investigation for general complex inhomogeneous terms, keeping however a compact $N_R$ spectrum.\par
The important conclusion of the calculations of the present Section is that the results presented in Sections 6 and 7 remain much more general than the very particular model that, for the sake of simplicity, was exposed there. Our scheme, although fine-tuned because we look for a compact heavy neutrino spectrum, allows for a wide range of parameters giving good results. 

\section{Comments on the compact heavy neutrino spectrum and on the level crossing region} \hspace*{\parindent}

We now go back to the case that we have studied in a quantitative detail, namely eqs. (\ref{6.1e})(\ref{6.2e}) with the conditions (\ref{6.3e})(\ref{6.4e}).\par
Before and around the level crossing region we have a rather or very compact heavy neutrino spectrum. For simplicity, we have assumed that the lightest heavy neutrino $N_{R_1}$ decays out-of-equilibrium and gives the main contribution to the important quantities relevant for baryogenesis : $\epsilon$, $\tilde{m}$ and $Y_{B}$.\par 
In such a fine-tuned situation, this can seem rather artificial. Actually, one should consider the contributions of all three heavy neutrinos, and therefore the contributions of all the CP-violation parameters $\epsilon_1$, $\epsilon_2$ and $\epsilon_3$, and the corresponding wash-out factors. Notice that there are studies in the literature that consider all these contributions. See, for example, the paper by E. Bertuzzo et al. \cite{DDBFN} and also, in a qualitative way, the work by Akhmedov et al. \cite{AFS}. However, to take into account the contributions of all three heavy neutrinos can present some subtleties. We have not dared for the moment to roughly add $Y_{B_1}$, $Y_{B_2}$ and $Y_{B_3}$. We simply expect that the possible contributions of all three heavy neutrinos will not strongly affect the results that we have found from $N_{R_1}$. An argument given below supports this hypothesis.\par

Let us now comment on the crossing region. As pointed out above, there is a level crossing in both cases : (1) $\tan \theta_s < 0$ and (2) $\tan \theta_s > 0$. This happens around $- C^L_{23} = C^L_{33} \simeq 3 \times 10^6\ GeV^{-1}$ for solution (1) and $C^L_{23} = - C^L_{33} \simeq 5 \times 10^5\ GeV^{-1}$ for solution (2). At some point in this region the heavy neutrinos $N_{R_1}$ and $N_{R_2}$ become degenerate.\par

But we want to make explicit more precisely what we understand by level crossing. What we mean is that the properties of the $N_{R_1}$ neutrino before the level crossing become (up to signs) those of the $N_{R_2}$ neutrino after the level crossing, and vice-versa, exchanging their effect on the absolute magnitude of the final quantities $\epsilon$, $\tilde{m}$ and $Y_{B}$. This can be seen easily by writing the effective Dirac neutrino mass that enters in formulas (\ref{4.2e}), (\ref{4.3e}), before and after the crossing region.\par

To illustrate what happens, we will take as an example solution (1). Similar features appear for solution (2). To be definite, we consider $N_{R_1}$ (the lightest neutrino) and $N_{R_2}$ (the next-to-lightest neutrino) before the crossing region, as we have computed in Section 7. Applying for $N_{R_2}$ naively the formulas (\ref{4.3e}), (\ref{4.6e}) and (\ref{4.8e}) making just the exchange $M_1 \leftrightarrow M_2$, let us give for solution (1) the quantities $\tilde{m}_1$, $\epsilon_1$, $Y_{B_1}$ and $\tilde{m}_2$, $\epsilon_2$, $Y_{B_2}$ at one point before the crossing region, for example for the value $C^L_{33} = 10^6\ GeV^{-1}$ and at one point after the crossing, for example for $C^L_{33} = 10^7\ GeV^{-1}$. Let us recall that the Dirac matrix (\ref{5.24e}) is completely fixed, but the redefined matrix (\ref{4.2e}) changes from point to point because it depends on the diagonalization of the matrix $M_R$ by (\ref{3.10e}). \par

One finds, {\it before the crossing}, for $C^L_{33} = 10^6\ GeV^{-1}$, the values for the heavy neutrino masses :
\beq
\label{10.1bise}
M_1 = 5.531 \times 10^9\ GeV \qquad M_2 = 1.017 \times 10^{10}\ GeV \qquad M_3 = 2.017 \times 10^{10}\ GeV
\eeq

\noindent and the hermitian matrix that enters in (\ref{4.6e}) and (\ref{4.3e}) : 

\beq
\label{10.1e}
\hat{m}_D^+\hat{m}_D \simeq  \left(
        \begin{array}{ccc}
        0.259 & 18.176 - 0.089i & -0.125 - 25.597i \\
       18.176 + 0.089i & 3352.06 & -0.00006 - 4720.59i \\
       -0.125 + 25.597i & -0.00006 + 4720.59i & 6647.84 \\
        \end{array}
        \right)\ GeV^2
\eeq        
and one gets therefore :
$$\tilde{m}_1 = 0.047\ eV \qquad \qquad \epsilon_1 = - 2.749 \times 10^{-6} \qquad \qquad  Y_{B_1} = 2.383 \times 10^{-11} \qquad \ \ \ $$
\beq
\label{10.2e}
\tilde{m}_2 = 329.586\ eV \qquad \qquad \epsilon_2 = -1.425 \times 10^{-10} \qquad \qquad Y_{B_2} = 4.245 \times 10^{-20}
\eeq

Notice here one point. The numbers obtained in (\ref{10.2e}) are very interesting in relation with the calculations done in Section 7 for the region before the level crossing. We have assumed there that $Y_B$ is dominated by the contribution of the lightest neutrino $Y_{B_1}$. We see indeed that, at least within these naive estimate, this is true as far as the consideration of the next-to-lightest neutrino is concerned.

Remember that we have adopted the level ordering convention (\ref{4.2bise}), that applies also after the crossing : $N_{R_1}$ is the lightest neutrino and $N_{R_2}$ the next-to-lightest neutrino.\par

One finds, {\it after the crossing}, for example for $C^L_{33} = 10^7\ GeV^{-1}$, the heavy neutrino masses :
\beq
\label{10.3bise}
M_1 = 2.011 \times 10^9\ GeV \qquad M_2 = 5.531 \times 10^9\ GeV \qquad M_3 = 1.020 \times 10^{11}\ GeV
\eeq

\noindent and the hermitian matrix that enters in (\ref{4.6e}) and (\ref{4.3e}) : 
\beq
\label{10.3e}
\hat{m}_D^+\hat{m}_D \simeq  \left(
        \begin{array}{ccc}
        193.27 & -5.937 - 0.020i & -0.00029 + 1376.69i \\
       -5.937 + 0.020i & 0.342 & -0.143 - 42.293i \\
       -0.00029 - 1376.69i & -0.143 + 42.293i & 9806.55 \\
        \end{array}
        \right)\ GeV^2
\eeq
and one gets therefore :
$$\tilde{m}_1 = 96.125\ eV \qquad \qquad \epsilon_1 = 8.023 \times 10^{-10} \qquad \qquad Y_{B_1} = -9.981 \times 10^{-19} \qquad \ \ \ $$
\beq
\label{10.4e}
\tilde{m}_2 = 0.062\ eV \qquad \qquad \epsilon_2 = 3.694 \times 10^{-6} \qquad \qquad Y_{B_2} = - 2.312 \times 10^{-11}
\eeq

The shifts in order of magnitude among the elements of the matrices before and after the crossing, (\ref{10.1e}) or (\ref{10.3e}), explain the strong differences (in magnitude and even in sign) of the relevant quantities in these two regions, (\ref{10.2e}) or (\ref{10.4e}). We observe a strong discontinuity for the lightest neutrino properties (and for the next-to-lightest ones) that happens going through the crossing region. Up to the sign of $\epsilon$ and therefore of $Y_B$, we see that after the crossing the lightest neutrino has very strong wash-out and very small $|\epsilon_1|$ and therefore $|Y_{B_1}|$, and that the opposite is true for the next-to-lightest heavy neutrino $N_{R_2}$. It is easy to examine this for the parameter $\tilde{m}_i$ (i = 1, 2), just by inspection of the matrix elements $(\hat{m}_D^+\hat{m}_D)_{ii}$ in (\ref{10.1e}) and (\ref{10.3e}). For $\epsilon_i$ it is a little more involved, but can be seen also by looking at the squares of the matrix elements of $\hat{m}_D^+\hat{m}_D$.\par

Let us now comment on the change of sign of $\epsilon_i$ and $Y_{B_i}$ before and after the level crossing region (\ref{10.2e}), (\ref{10.4e}). For solution (1), that we discuss here, the sign of the inhomogeneous term (\ref{6.3e}) $- C^L_{23} = C^L_{33} > 0$ gives $\epsilon_i < 0$ and $Y_{B_i} > 0$ before the crossing, and $\epsilon_i > 0$ and $Y_{B_i} < 0$ after the crossing. We have realized that if one changes the sign of (\ref{6.3e}), i.e. $- C^L_{23} = C^L_{33} < 0$, then one has the opposite :  $\epsilon_i > 0$ and $Y_{B_i} < 0$ before the crossing, and $\epsilon_i < 0$ and $Y_{B_i} > 0$ after the crossing. Adopting this latter sign for $C^L_{23}, C^L_{33}$, nothing essential changes for the heavy neutrino spectrum and for $\Delta m_s^2$ and $\Delta m_a^2$. The same considerations apply to solution (2) using (\ref{6.4e}) and changing its sign.\par
Our conclusion is that, provided $N_{R_1}$ and $N_{R_2}$ are close enough in mass, one can have the right order of magnitude and sign for $Y_B$ before and after the crossing. This happens only at the price of changing the sign of our single free real parameter $- C^L_{23} = C^L_{33}$.\par 
An interesting conclusion is that, after the level crossing, the second-to-lightest heavy neutrino $N_{R_2}$ dominates. The possibility of next-to-lightest neutrino dominance has been extensively studied recently by S. Antush et al. \cite{ADBJK}. 

\par \vskip 1.0 truecm

\section{Open problems within the present approach} \hspace*{\parindent}

There are a number of problems to face and study within the present approach. Let us make an incomplete list :

\par \vskip 0.3 truecm

(i) There is the possibility that more than one heavy right-handed neutrino decays out of equilibrium, contributing to the leptogenesis, a point that, in particular, has been suggested rather clearly in ref. \cite{AFS}. If we guess a temperature $T \simeq 10^{11}\ GeV$ below which all heavy neutrinos decay out-of-equilibrium, then not only the lightest $N_{R_1}$ decays out-of equilibrium after the level crossing, but also $N_{R_2}$ is in the same situation. On the other hand, the heavy neutrino spectrum being rather compact, the natural thing to do would be to consider the contributions to $Y_B$ of all three heavy neutrinos $N_{R_1}$, $N_{R_2}$ and $N_{R_3}$, i.e. to compute $\epsilon_1$, $\epsilon_2$ and $\epsilon_3$ and the relevant washout factors.\par
For the moment we just expect that the consideration of the three neutrinos will not spoil the good features of the calculations of the present paper, that take only into account the lightest neutrino $N_{R_1}$before the crossing.\par
We have given an argument going in this sense in Section 10 where we have seen that, before the level crossing, the contribution of the next-to-lightest neutrino $N_{R_2}$ is negligible compared to the one of the lightest one $N_{R_1}$. 

\par \vskip 0.2 truecm

(ii) One should take into account also the level crossing region and therefore the finite width of the right-handed neutrinos, as well as the delicate question of their interference. These problems have been treated in great detail by A. Pilaftsis and collaborators \cite{PILAFTSIS} that, to be complete, need to be adopted within our approach.

\par \vskip 0.2 truecm

(iii) The flavor effects, thoroughly studied by A. Abada et al. \cite{ABADA} are also a delicate question to study in this region of compact right-handed neutrino spectrum, and this should also be performed.  

\par \vskip 0.2 truecm

(iv) It would be worth to study the more general case for CP violation outlined in Section 9, and make a detailed scan of the results in the case of general complex inhomogeneous terms - or equivalently general light neutrino Majorana phases -, with the constraint of having a compact heavy neutrino spectrum.

\par \vskip 0.2 truecm

(v) It could be that the homogeneous equations (\ref{5.1e}) correspond to some symmetry, the inhomogeneous terms (that we have introduced to get a large enough CP violation $\epsilon_1$) being a breaking of this symmetry, a possibility that would be interesting. 

\par \vskip 1.0 truecm

\section{Conclusions} \hspace*{\parindent} 

Our demand of a compact $N_R$ spectrum, and of an approximate quark-lepton symmetry implying a hierarchical spectrum for the Dirac neutrino masses with a similar structure between $V^L$ and the CKM mixing matrix, brings to a scenario where the lepton asymmetry comes out naturally, producing the required order of magnitude for the baryon asymmetry $Y_{B} \sim O(10^{-10})$. We have assumed and justified that $Y_B$ is dominated by the contribution of the lightest neutrino $N_{R_1}$.\par
In this way, not only one can get a good magnitude for $Y_{B}$, but as a natural consequence there are also a number of other strong points in this approach.

\par \vskip 0.2 truecm

We get two possible solutions with a normal hierarchical light neutrino mass spectrum and an absolute scale, i.e. the lightest neutrino mass $m_1$ must be non vanishing.

\par \vskip 0.2 truecm

The light neutrino squared mass differences $\Delta m_s^2$ and $\Delta m_a^2$ are very stable and consistent with the data.

\par \vskip 0.2 truecm

There are three CP-violating phases in the whole approach, the phase $\delta_L$ of the $V^L$ unitary matrix, and the light neutrino Majorana phases $\alpha$ and $\beta$. We take $\delta_L$ to be of the order of the Kobayashi-Maskawa phase $\delta_{KM}$.

\par \vskip 0.2 truecm

We have thoroughly studied in a quantitative way a particular case in which all CP violating effects are computed in terms of $\delta_L$, in particular $\epsilon_1$ and $Y_{B_1}$. It is interesting that one can get a baryon asymmetry of the right order of magnitude taking $\delta_L \simeq \delta_{KM}$. Of course, this result is not obtained in the Standard Model, but in a New Physics scheme under particular assumptions : Baryogenesis via Leptogenesis, $SO(10)$ Grand Unification, approximate quark-lepton symmetry and compact heavy $N_R$ spectrum. In the limit $\delta_L \to 0$, one gets indeed $\epsilon_1 \to 0$ and $Y_{B_1} \to 0$. 

\par \vskip 0.2 truecm

The $\nu_e$ mass, bounded by tritium $\beta$ decay, is of the order of few times $10^{-3}\ eV$.

\par \vskip 0.2 truecm

The sum $\sum_i m_{\nu_i}$ satisfies the cosmological bounds, with a value rather close to the present upper limits.

\par \vskip 0.2 truecm

Let us emphasize that $\delta_L$ induces also small CP violating corrections to the light neutrino Majorana phases, that turn out to be naturally close to $\alpha = {\pi \over 2}, \beta = {\pi \over 2} \ \textrm{or} \ 0$. The effective neutrino mass, relevant for neutrinoless double beta decay, comes out to be rather small, of the order of $10^{-3}\ eV$, because of strong cancellations due to the Majorana phases.

\par \vskip 0.2 truecm

In the region of quasi-degeneracy, the heaviest $N_R$ has a mass of the order $1.5 \times 10^{10}\ GeV$, roughly consistent with the expected scale of $B-L$ symmetry breaking, so that $SO(10)$ breaks down to the Pati-Salam group $SU(4) \times SU(2) \times SU(2)$ at the expected natural intermediate scale.

\par \vskip 0.2 truecm

We expose also an example in which the phase of the Dirac neutrino mass matrix $\delta_L$ and the Majorana phases $\alpha$, $\beta$ are independent, providing an efficient generation of baryon asymmetry.

\par \vskip 1.0 truecm

\noindent {\large \bf Acknowledgements}

\par \vskip 0.5 truecm

This work has been supported in part by the EU Contract No. MRTN-CT-2006-035482, FLAVIAnet. We are indebted to A. Abada and F.-X. Josse-Michaux for discussions in the early stage of this work.

\par \vskip 1.0 truecm

{\large \bf Appendix}

\par \vskip 1.0 truecm

In this Appendix we demonstrate the approximate formula (\ref{1.13e}) for $\Delta m_a^2$ :

\par \vskip 0.5 truecm

\qquad \qquad $\Delta m_a^2 = |m_3|^2 - \textrm{cos}^2\theta_s\ |m_2|^2 - \textrm{sin}^2 \theta_s\ |m_1|^2$ \qquad \qquad \qquad \qquad \qquad \qquad (A.1)

\par \vskip 0.5 truecm

\noindent The mass eigenstates read :

\par \vskip 0.5 truecm

\qquad \qquad $|\nu_1>\ = c_s\ |\nu_e> -\ {s_s \over \sqrt{2}}\ |\nu_{\mu}> +\ {s_s \over \sqrt{2}}\ |\nu_{\tau}>$

\par \vskip 0.2 truecm

\qquad \qquad $|\nu_2>\ = s_s\ |\nu_e> +\ {c_s \over \sqrt{2}}\ |\nu_{\mu}> -\ {c_s \over \sqrt{2}}\ |\nu_{\tau}>$

\par \vskip 0.2 truecm

\qquad \qquad $|\nu_3>\ = {1 \over \sqrt{2}}\ |\nu_{\mu}> +\ {1 \over \sqrt{2}}\ |\nu_{\tau}>$ \qquad \qquad \qquad \qquad \qquad \qquad \qquad \qquad (A.2)

\par \vskip 0.5 truecm

\noindent and therefore the $\mu$-neutrino state is given, in terms of the mass eigenstates :

\par \vskip 0.5 truecm

\qquad \qquad $|\nu_{\mu}>\ = -\ {s_s \over \sqrt{2}}\ |\nu_1> +\ {c_s \over \sqrt{2}}\ |\nu_2> +\ {1 \over \sqrt{2}}\ |\nu_3>$ \qquad \qquad \qquad \qquad \qquad (A.3)
 
\par \vskip 0.5 truecm
 
\noindent that evolves in time according to :
 
 \par \vskip 0.5 truecm
 
\noindent $|\nu_{\mu}(t)>\ = -\ {s_s \over \sqrt{2}}\ e^{(-ip-i{m_1^2 \over {2p}}) t}|\nu_1> +\ {c_s \over \sqrt{2}}\ e^{(-ip-i{m_2^2 \over {2p}})t} |\nu_2> +\ {1 \over \sqrt{2}}\ e^{(-ip-i{m_3^2 \over {2p}})t} |\nu_3>$\par 
$= e^{(-ip-i{m_3^2 \over {2p}})t} \left[-\ {s_s \over \sqrt{2}}\ e^{i{({{m_3^2-m_1^2} \over {2p}})t}}|\nu_1> +\ {c_s \over \sqrt{2}}\ e^{i{({{m_3^2-m_2^2} \over {2p}})t}} |\nu_2> +\ {1 \over \sqrt{2}}\ |\nu_3>\right]$ \qquad \qquad (A.4)

\par \vskip 0.5 truecm
 
\noindent and, from (A.2), we can write the scalar products :

\par \vskip 1.0 truecm

\noindent $e^{(ip+i{m_3^2 \over {2p}})t} <\nu_e|\nu_{\mu}(t)>$\par
$= \left[-\ {s_s \over \sqrt{2}}\ e^{i{({{m_3^2-m_1^2} \over {2p}})t}}<\nu_e|\nu_1> +\ {c_s \over \sqrt{2}}\ e^{i{({{m_3^2-m_2^2} \over {2p}})t}} <\nu_e|\nu_2> +\ {1 \over \sqrt{2}}\ <\nu_e|\nu_3>\right]$\par
$= {s_s c_s \over \sqrt{2}} \left[-\ e^{i{({{m_3^2-m_1^2} \over {2p}})t}} +\ e^{i{({{m_3^2-m_2^2} \over {2p}})t}}\right] = {s_s c_s \over \sqrt{2}} e^{i{({{m_3^2-m_2^2} \over {2p}})t}} \left[-\ 1 +\ e^{i{({{m_2^2-m_1^2} \over {2p}})t}}\right]$ \qquad \qquad (A.5)

\par \vskip 0.5 truecm

\noindent $e^{(ip+i{m_3^2 \over {2p}})t} <\nu_{\mu}|\nu_{\mu}(t)>$\par
$= \left[-\ {s_s \over \sqrt{2}}\ e^{i{({{m_3^2-m_1^2} \over {2p}})t}}<\nu_{\mu}|\nu_1> +\ {c_s \over \sqrt{2}}\ e^{i{({{m_3^2-m_2^2} \over {2p}})t}} <\nu_{\mu}|\nu_2> +\ {1 \over \sqrt{2}}\ <\nu_{\mu}|\nu_3>\right]$\par
$= {1 \over 2} \left[s_s^2\ e^{i{({{m_3^2-m_1^2} \over {2p}})t}} +\ c_s^2\ e^{i{({{m_3^2-m_2^2} \over {2p}})t}} + 1\right]$ \qquad \qquad \qquad \qquad \qquad \qquad \qquad  \qquad \qquad (A.6)

\par \vskip 0.5 truecm

\noindent $e^{(ip+i{m_3^2 \over {2p}})t} <\nu_{\tau}|\nu_{\mu}(t)>$\par
$= \left[-\ {s_s \over \sqrt{2}}\ e^{i{({{m_3^2-m_1^2} \over {2p}})t}}<\nu_{\tau}|\nu_1> +\ {c_s \over \sqrt{2}}\ e^{i{({{m_3^2-m_2^2} \over {2p}})t}} <\nu_{\tau}|\nu_2> +\ {1 \over \sqrt{2}}\ <\nu_{\tau}|\nu_3>\right]$\par
$= {1 \over 2} \left[-\ s_s^2\ e^{i{({{m_3^2-m_1^2} \over {2p}})t}} -\ c_s^2\ e^{i{({{m_3^2-m_2^2} \over {2p}})t}} + 1 \right]$ \qquad \qquad \qquad \qquad \qquad \qquad \qquad \qquad (A.7)

\par \vskip 0.5 truecm

\noindent Denoting 

\qquad \qquad $\alpha_{ji} = {(m_j^2-m_i^2)t \over {2p}}$ \qquad \qquad \qquad \qquad \qquad \qquad \qquad \qquad \qquad \qquad \qquad \qquad (A.8)

\par \vskip 0.5 truecm

\noindent One obtains, from (A.7) and (A.8) and the relation

\par \vskip 0.5 truecm

\qquad \qquad $\alpha_{31} - \alpha_{32} = \alpha_{21}$  \qquad \qquad \qquad \qquad \qquad \qquad \qquad \qquad \qquad \qquad \qquad (A.9)

\par \vskip 0.5 truecm

\noindent the following probabilities : 

\par \vskip 0.5 truecm

$|<\nu_e|\nu_{\mu}(t)>|^2\ = s_s^2c_s^2\ \left[1 - \cos(\alpha_{21})\right]$\qquad \qquad \qquad  \qquad  \qquad  \qquad  \qquad  \qquad (A.10)

\par \vskip 0.5 truecm

$|<\nu_\mu|\nu_{\mu}(t)>|^2\ = {1 \over 4} \left[s_s^4 + c_s^4 + 1 + 2s_s^2c_s^2\ \cos(\alpha_{21}) + 2 s_s^2\ \cos(\alpha_{31}) + 2 c_s^2\ \cos(\alpha_{32}) \right] \ \ (A.11)$

\par \vskip 0.5 truecm

$|<\nu_{\tau}|\nu_{\mu}(t)>|^2\ = {1 \over 4} \left[s_s^4 + c_s^4 + 1 + 2s_s^2c_s^2\ \cos(\alpha_{21}) - 2 s_s^2\ \cos(\alpha_{31}) - 2 c_s^2\ \cos(\alpha_{32}) \right] \ \ (A.12)$

\par \vskip 0.5 truecm

\noindent and one obtains, as expected :

\par \vskip 0.5 truecm 

$|<\nu_e|\nu_{\mu}(t)>|^2\ +\ |<\nu_\mu|\nu_{\mu}(t)>|^2\ +\ |<\nu_{\tau}|\nu_{\mu}(t)>|^2\ = 1 \qquad \qquad \qquad \qquad (A.13)$

\par \vskip 0.5 truecm 

\noindent Performing an expansion in powers of $\alpha_{ji} = {(m_j^2-m_i^2)t \over {2p}}$, one finds

\par \vskip 0.5 truecm 

$|<\nu_e|\nu_{\mu}(t)>|^2\ \simeq {s_s^2c_s^2 \over 2} \left[{(m_2^2-m_1^2)t \over {2p}}\right]^2$\qquad \qquad \qquad  \qquad  \qquad  \qquad  \qquad  \qquad  \qquad (A.14)

\par \vskip 0.5 truecm

$|<\nu_\mu|\nu_{\mu}(t)>|^2\ \simeq 1 - {s_s^2c_s^2 \over 4} \left[{(m_2^2-m_1^2)t \over {2p}}\right]^2 - {s_s^2 \over 4} \left[{(m_2^2-m_1^2)t \over {2p}}\right]^2 - {c_s^2 \over 4} \left[{(m_3^2-m_2^2)t \over {2p}}\right]^2$ \qquad  \qquad (A.15)

\par \vskip 0.5 truecm

$|<\nu_\tau|\nu_{\mu}(t)>|^2\ \simeq - {s_s^2c_s^2 \over 4} \left[{(m_2^2-m_1^2)t \over {2p}}\right]^2 + {s_s^2 \over 4} \left[{(m_2^2-m_1^2)t \over {2p}}\right]^2 + {c_s^2 \over 4} \left[{(m_3^2-m_2^2)t \over {2p}}\right]^2$ \qquad  \qquad (A.16)

\par \vskip 0.5 truecm

\noindent with :

\par \vskip 0.5 truecm 

$|<\nu_e|\nu_{\mu}(t)>|^2\ +\ |<\nu_\mu|\nu_{\mu}(t)>|^2\ +\ |<\nu_{\tau}|\nu_{\mu}(t)>|^2\ \simeq 1 \qquad \qquad \qquad \qquad (A.17)$

\par \vskip 0.5 truecm

\noindent The last terms in the r.h.s. of (A.16) and (A.17) reads :

\par \vskip 0.5 truecm

${s_s^2 \over 4} \left[{(m_2^2-m_1^2)t \over {2p}}\right]^2 + {c_s^2 \over 4} \left[{(m_3^2-m_2^2)t \over {2p}}\right]^2 = {1 \over 4} [s_s^2(m_3^2-m_1^2)^2+c_s^2(m_3^2-m_2^2)^2] \left({t \over {2p}}\right)^2$  \qquad \qquad (A.18)

\par \vskip 0.5 truecm

\noindent and, for $m_1^2, m_2^2 << m_3^2$, the bracket in (A.18) becomes

\par \vskip 0.5 truecm

$[s_s^2(m_3^2-m_1^2)^2+c_s^2(m_3^2-m_2^2)^2] \simeq [m_3^2-(s_s^2m_1^2+c_s^2 m_2^2)]^2$  \qquad \qquad  \qquad  \qquad (A.19)

\par \vskip 0.5 truecm

\noindent and therefore, formulas (A.14)-(A.16) can be approximated by

\par \vskip 0.5 truecm 

$|<\nu_e|\nu_{\mu}(t)>|^2\ \simeq {s_s^2c_s^2 \over 2} \left[{(m_2^2-m_1^2)t \over {2p}}\right]^2$\qquad \qquad \qquad  \qquad  \qquad  \qquad  \qquad \qquad (A.20)

\par \vskip 0.5 truecm

$|<\nu_\mu|\nu_{\mu}(t)>|^2\ \simeq 1 - {s_s^2c_s^2 \over 4} \left[{(m_2^2-m_1^2)t \over {2p}}\right]^2 - {1 \over 4} \left[{(m_3^2-m_x^2)t \over {2p}}\right]^2$ \qquad  \qquad  \qquad \qquad (A.21)

\par \vskip 0.5 truecm

$|<\nu_\tau|\nu_{\mu}(t)>|^2\ \simeq - {s_s^2c_s^2 \over 4} \left[{(m_2^2-m_1^2)t \over {2p}}\right]^2 + {1 \over 4} \left[{(m_3^2-m_x^2)t \over {2p}}\right]^2$ \qquad  \qquad  \qquad \qquad (A.22)

\par \vskip 0.5 truecm

\noindent that satisfies (A.17) and where

\par \vskip 0.5 truecm

\qquad \qquad $m_x^2 = s_s^2m_1^2+c_s^2 m_2^2$ \qquad \qquad \qquad  \qquad  \qquad  \qquad  \qquad  \qquad \qquad \qquad  (A.23)

\par \vskip 0.5 truecm

\noindent Therefore the formula (A.1) or (\ref{1.13e}) follows. 

\par 

\end{document}